\documentclass{WileyMSP-template}

\usepackage{xcolor} % colored text
\usepackage[version=3]{mhchem} % Formula subscripts using \ce{}
\usepackage{csquotes}
\usepackage{url} % cite websites
\usepackage{xr} % reference figures in the SI.tex
\usepackage{appendix} % to create the environment for SI figures

\begin{document}

\pagestyle{plain}

%%%%%%%%%%%%%%%%%%%%%%%%%%%%%%%%%%%%%%%%%%%%%%%%%%%%%%%%%%%%%%%%%%%%%
%% Title
%%%%%%%%%%%%%%%%%%%%%%%%%%%%%%%%%%%%%%%%%%%%%%%%%%%%%%%%%%%%%%%%%%%%%

\title{In situ growth and magnetic characterization of Cr Chloride monolayers}
\maketitle
\vskip 0.2in

%%%%%%%%%%%%%%%%%%%%%%%%%%%%%%%%%%%%%%%%%%%%%%%%%%%%%%%%%%%%%%%%%%%%%
%% Authors and affiliations
%%%%%%%%%%%%%%%%%%%%%%%%%%%%%%%%%%%%%%%%%%%%%%%%%%%%%%%%%%%%%%%%%%%%%

\author{Giuseppe Buccoliero,} 
\author{Pamella Vasconcelos Borges Pinho,}
\author{Marli dos Reis Cantarino,}
\author{Francesco Rosa,}
\author{Nicholas B. Brookes,}
\author{Roberto Sant}*

\begin{affiliations}
\vskip 0.2in
G. Buccoliero, Dr. P.V.B. Pinho, Dr. M.R. Cantarino, Dr. N.B. Brookes, Dr. R. Sant\\
ESRF, The European Synchrotron, 71 Avenue des Martyrs, CS40220, 38043 Grenoble Cedex 9, France
\vskip 0.1in
Dr. P.V.B. Pinho\\\
Université Paris-Saclay, CEA, Service de Recherche en Corrosion et Comportement des Matériaux, F-91191 Gif-sur-Yvette, France
\vskip 0.1in
Dr. R. Sant, F. Rosa\\
Dipartimento di Fisica, Politecnico di Milano, Piazza Leonardo da Vinci 32, I-20133 Milano, Italy.\\
Email Address: roberto.sant@polimi.it
\end{affiliations}

%%%%%%%%%%%%%%%%%%%%%%%%%%%%%%%%%%%%%%%%%%%%%%%%%%%%%%%%%%%%%%%%%%%%%
%% Keywords
%%%%%%%%%%%%%%%%%%%%%%%%%%%%%%%%%%%%%%%%%%%%%%%%%%%%%%%%%%%%%%%%%%%%%

\vskip 0.2in

%%%%%%%%%%%%%%%%%%%%%%%%%%%%%%%%%%%%%%%%%%%%%%%%%%%%%%%%%%%%%%%%%%%%%
%% Abstract
%%%%%%%%%%%%%%%%%%%%%%%%%%%%%%%%%%%%%%%%%%%%%%%%%%%%%%%%%%%%%%%%%%%%%

\vskip 0.2in
\begin{abstract}
Monolayer Chromium Dihalides and Trihalides materials can be grown on a variety of substrates by molecular beam epitaxy regardless of the lattice mismatch thanks to the van der Waals epitaxy.
In this work, we studied the magnetic nature of Cr Chloride monolayers grown on Au(111), Ni(111) and graphene-passivated Ni(111) from the evaporation in ultra-high vacuum of the same halide precursor.
Structural, morphological and magnetic characterizations were conducted \textit{in situ} by low energy electron diffraction (LEED), scanning tunneling microscopy (STM) and X-ray magnetic circular dichroism (XMCD).
Owing to opposite chemical behaviour, Au(111) and Ni(111) promote the formation of two different valence compounds, \textit{i.e.} CrCl$_3$ and CrCl$_2$, showing distinct magnetic properties at 4 K. 
When graphene is used to passivate the Ni(111) surface, the formation of CrCl$_3$ becomes allowed also on this substrate. The coexistence of CrCl$_3$ and CrCl$_2$, both showing few nm lateral size and super-paramagnetic properties, is demonstrated by XMCD spectra displaying two dichroic peaks at the characteristic Cr$^{3+}$ and Cr$^{2+}$ energies.
\textit{Site-selective} magnetization measurements performed with the photon energy tuned on the two absorption edges show reversed magnetization of some of the CrCl$_2$ islands with respect to the CrCl$_3$ domains, which is interpreted in terms of magnetic frustration.
\end{abstract}

%%%%%%%%%%%%%%%%%%%%%%%%%%%%%%%%%%%%%%%%%%%%%%%%%%%%%%%%%%%%%%%%%%%%%
%% Introduction
%%%%%%%%%%%%%%%%%%%%%%%%%%%%%%%%%%%%%%%%%%%%%%%%%%%%%%%%%%%%%%%%%%%%%

\section{Introduction}
Chromium Dihalides (CrX$_2$, X=Cl,Br,I) and Trihalides (CrX$_3$) van der Waals (vdW) materials exhibit intriguing layered magnetic order with different spin arrangements within the layer units. These configurations are governed by the competition between the antiferromagnetic (AFM) direct exchange among nearest-neighbours metal atoms and the ferromagnetic (FM) super-exchange between second neighbours mediated by halogen \textit{p} states \cite{kulish,mcguire_review}.
In CrX$_2$ compounds the metal cations are arranged in triangular nets and the materials develop AFM properties. When in-plane Jahn-Teller distortions are active, edge-sharing octahedra form ribbons that are characterized by helimagnetic spin order \cite{schneeloch,liu}.
On the other hand, in CrX$_3$, the transition metal cations form honeycomb lattices where the in-plane interactions are of FM type \cite{mcguire_CrI3, tian}. Layer stacking gives rise in some cases to AFM interlayer coupling and overall AFM behavior, as for instance in CrCl$_3$ \cite{mcguire_CrCl3}.

Exfoliated single layers (SL) halides can be used as building blocks to incorporate magnetism into functional vdW two-dimensional (2D) heterostructures. The vdW nature of the interface minimizes effects such as hybridization, charge doping and other substrate-driven crystalline effects. However, magnetic proximity might result in varying the exchange coupling across the interface \cite{gibertini}.
To build up devices with sharp and clean interfaces and scale up fabrication, bottom-up growth methods based on epitaxy are instead required.
The push to synthesize single layer epitaxial halides via molecular beam epitaxy (MBE) has been driven by the simplicity of the process, which consists for most cases \cite{bedoya} in the evaporation under ultra-high-vacuum (UHV) of the very same stoichiometric powder compound.
In this context, a single crystal substrate provides a suitable platform where low dimensional magnetism can be examined. Inert and reactive, as well as magnetic and non-magnetic substrate surfaces can be tested for this purpose, while suitable recipes have to be developed to control the growth in favor of the formation of a single targeted species. Nevertheless, the synthesis of multiple coexisting phases can be desirable to build 2D patterned lateral heterostructures \cite{li}.
The freestanding properties of epitaxial 2D layers might be drastically altered when grown on tridimensional non-vdW substrates due to hybridization via dangling bonds and charge transfer. In other cases, magnetic proximity effects can be engineered on purpose to tune the pristine properties. The substrate choice must always be done considering the chemical affinity towards the precursors, since it might be detrimental to the stabilization of the desired phases.
In similar cases, the substrate can be passivated by other 2D materials, such as graphene or boron nitride, to hinder the reactivity \cite{lodesani,sant_WSe2}.
From a technological point of view, spin injection into graphene by proximity effect from the growth substrate is an interesting and developing field \cite{yang}. 2D graphene in fact is considered as one of the model systems for nanospintronics applications due to its high electron mobility and long spin relaxation length \cite{schwierz}.

In this work we have studied the growth products of CrCl$_3$ evaporation in UHV on different types of metallic single crystal substrates.
CrCl$_3$ is a vdW material consisting of FM layer units having in-plane long range magnetic order below 17 K. In the magnetic phase, bulk compounds have rhombohedral \textit{R-3} space group symmetry and the layers are stacked with AFM spin order \cite{mcguire_CrCl3}, while the SL remains intrinsically FM \cite{bedoya}. The individual layers of CrCl$_3$ are constituted by a Cr honeycomb net where Cr$^{3+}$ cations in S = 3/2 spin state are octahedrally coordinated by six Cl$^{-1}$ anions. Each Cl atom is shared between two Cr atoms and mediates the intra-layer FM super-exchange interaction. 
Early experimental works on bulk compounds show instead that crystalline CrCl$_2$ has an orthorhombic unit cell and \textit{Pnnm} space group symmetry \cite{hermann}. Similarly to CrI$_2$, Cr atoms dimerize as a result of lattice deformation (Cr atoms are surrounded by a distorted Cl octahedron), forming chains along which spins have helical configuration \cite{schneeloch}. However, a study on the magnetic properties of these species grown as epitaxial thin films is still lacking.

Herein, we will describe the different nature of Chloride monolayers obtained by the evaporation in UHV of a CrCl$_3$ powder precursor onto different metallic single crystal surfaces, \textit{i.e.} a non-magnetic and inert Au(111) and a magnetic but chemically reactive Ni(111) substrates. A comprehensive study linking the spectroscopic information and the magnetic properties to structural and morphological characterizations is in fact still missing in the literature, whereas it is of paramount importance to choose and manipulate materials in view of device applications.  
The as grown monolayers are characterized by a combination of \textit{in situ} techniques such as Low Energy Electron Diffraction (LEED), Scanning Tunneling Microscopy (STM), X-ray Absorption Spectroscopy (XAS) and X-ray Magnetic Circular Dichroism (XMCD) in order to obtain complementary structural, morphological and spectroscopical information. 
We will show that Au(111) and Ni(111) promote the formation of Chloride species with different valence, \textit{i.e.} CrCl$_3$ and CrCl$_2$, showing also distinct magnetic behaviors.
In order to prevent the reduction of the adsorbed molecules from Cr$^{3+}$ to Cr$^{2+}$, the Ni substrate was passivated by a single layer of graphene (SLG), and results from samples synthesized with different processes are compared. 
SLG favors the partial formation of CrCl$_3$ domains in addition to CrCl$_2$ ones, both showing few nm lateral size and super-paramagnetic (SPM) features.
The coexistence of the CrCl$_2$ and CrCl$_3$ monolayers is evidenced by XMCD spectra showing two dichroic peaks at the characteristic Cr$^{2+}$ and Cr$^{3+}$ absorption energies, and by site-selective magnetization measurements performed by tuning the probing photon energy to the two specific edges.
This method - allowed by the tunability of the synchrotron undulator source - revealed the antiparallel magnetization of some of the CrCl$_2$ domains with respect to the CrCl$_3$ ones and the applied magnetic field, which led to an inverted magnetization loop. The result is interpreted in terms of competing magnetic orders between the two different species.

%%%%%%%%%%%%%%%%%%%%%%%%%%%%%%%%%%%%%%%%%%%%%%%%%%%%%%%%%%%%%%%%%%%%%
%% Main Text 
%% on Au111
%%%%%%%%%%%%%%%%%%%%%%%%%%%%%%%%%%%%%%%%%%%%%%%%%%%%%%%%%%%%%%%%%%%%%

\section{Results and Discussion}
SL Cr Chlorides have been grown on Au(111), Ni(111) and graphene-passivated Ni(111), by evaporation from a Knudsen cell of a anhydrous CrCl$_3$ powder compound, as described in the \textit{Experimental Section}. Morphology and structure of the as grown epitaxial 2D materials were firstly characterized \textit{in situ} by LEED and STM in the same MBE setup and then transferred in UHV into the high-field-magnet (HFM) chamber connected with the synchrotron beam for the spectroscopic and magnetic characterization (XAS and XMCD). To assist in interpreting the acquired spectra, ligand field multiplet (LFM) calculations \cite{degroot} were performed using the \textit{Quanty} code \cite{haverkort}. Hereafter, we describe the results obtained for each type of sample.

\subsection{CrCl$_3$/Au(111)}
\begin{figure}
    \centering
    \includegraphics[scale=0.8]{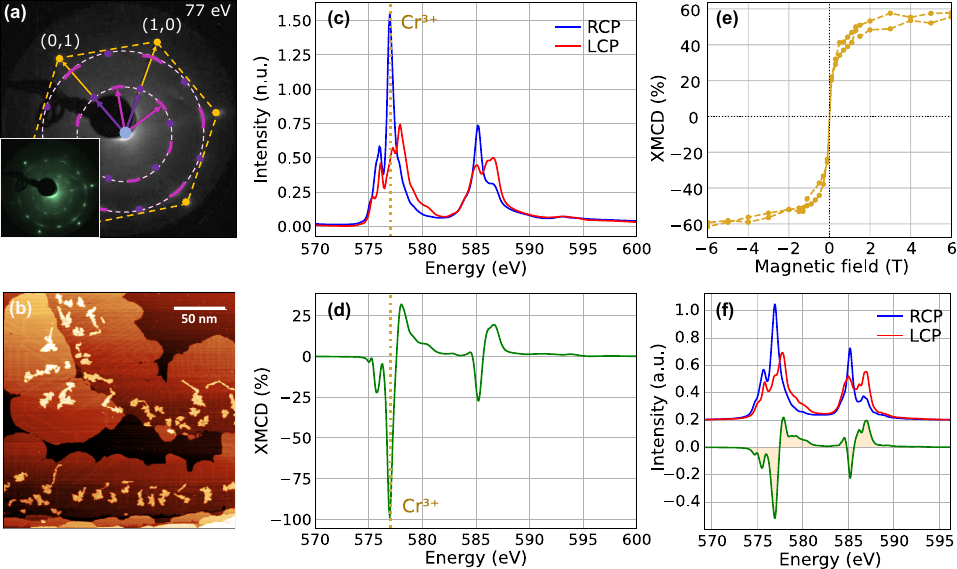}
    \caption{
    \textbf{CrCl$_3$/Au(111).} \textbf{(a)} LEED pattern acquired at 77 eV. Yellow spots and vectors correspond to substrate reflections and unit cell, while pink and violet ones indicate (2$\times$2) and (2$\times$2)R30$^{\circ}$ CrCl$_3$ reconstructions; the white dashed circumferences highlight the ring signals crossing the overlayer spots; \textit{inset}: same LEED image without guidelines. \textbf{(b)} STM image (225$\times$225 nm$^2$) showing SL CrCl$_3$ on Au(111) surface; second layer clusters are also visible on top of the CrCl$_3$ domains. \textbf{(c-d)} XAS and XMCD spectra taken around Cr L$_{3,2}$ edges at 4 K, under 9 T magnetic field applied along the beam direction, and in NI; the RCP and LCP XAS intensities are normalized to the average; the XMCD intensity is expressed in percentage units relative to the average XAS. \textbf{(e)} Magnetization curves measured with 577.0 eV photon energy (Cr$^{3+}$ edge, indicated by the dashed line in (c) and (d)), at 4 K and in NI. \textbf{(f)} Simulations of the experimental XAS and XMCD spectra shown in (c) and (d).
    }
    \label{fig:Au}
\end{figure}

A clean herringbone reconstructed Au(111) single crystal surface was exposed to a CrCl$_3$ flux at 150$^{\circ}$C, as described in the \textit{Experimental Section}. 
The LEED pattern in Figure \ref{fig:Au}a reveals that the overlayer grows oriented along two epitaxial directions, \textit{i.e.} aligned and 30$^{\circ}$ off the substrate crystallographic axes, forming commensurate (2$\times$2) and incommensurate (2$\times$2)R30$^{\circ}$ reconstructions, the second characterized by a significantly larger mosaic spread. 
The estimated overlayer surface lattice constant is $\sim$5.8 \AA, close to the in-plane lattice parameter of bulk CrCl$_3$ (5.96 \AA) \cite{mcguire_CrCl3}. 
The two rings crossing the overlayer spots indicate poor orientational alignment of the as-grown domains compared to the substrate and point to a weak interface bonding. This is a typical signature of the vdW epitaxy, which allows the growth of 2D films on top of the substrate also in the presence of a significant lattice mismatch \cite{koma,walsh}.

The STM image in Figure \ref{fig:Au}b shows the morphology of the overlayer. The domains consist of patches of several hundreds of nm lateral size and irregular contours. 
Brighter features, probably second layer clusters, are also visible.
No periodic intensity modulation was observed on the surface, \textit{e.g.} moiré superstructures or other kinds of patterning, but flat areas which are reminiscent of the denominated \enquote{pristine phase} recorded for CrI$_3$ grown on Au(111). \cite{li_CrI3}.
The apparent height of $\sim$3 \AA (see Figure \ref{fig:z_profile}b in the \textit{Supporting Information}) is compatible with the thickness of Cl-Cr-Cl trilayers bonded to tridimensional substrates with no vdW gap.

After the preliminary structural and morphological characterization, the sample was transferred into the HFM chamber for the spectroscopy measurements with synchrotron X-rays. 
Figure \ref{fig:Au}c shows the XAS spectra acquired with right-circular-polarized (RCP) and left-circular-polarized (LCP) X-rays around the Cr L$_{3,2}$ edges at 4 K, under 9 T applied magnetic field and in normal incidence (NI) geometry.
The spectra can be divided in the spin-orbit split L$_3$ ($\sim$577.0 eV) and L$_2$ ($\sim$585.5 eV) parts. 
An assessable XMCD signal is associated to element-specific magnetic polarization. On this sample, the XMCD intensity achieves the 98\% of the average XAS intensity at the Cr L$_3$ edge under an applied magnetic field of 9 T (Figure \ref{fig:Au}d). 
The spectra replicates the results obtained by \textit{Bedoya-Pinto et al.} \cite{bedoya} and well reproduces the XAS lineshape of a commercial bulk CrCl$_3$ reference \cite{hqgraphene} measured in the same setup, as reported in the \textit{Supporting Information} (Figure \ref{fig:xmcd_bulk}). 
This evidence confirms the hypothesis that the as grown monolayers are CrCl$_3$ species and that the main peak position (577.0 eV) corresponds to a Cr$^{3+}$ valence state.

Magnetization cycles have been acquired by measuring the XMCD intensity as a function of the magnetic field varied between -6T and 6T, using the photon energy corresponding to the XMCD peak. 
The result, shown in Figure \ref{fig:Au}e, is a closed S-shaped loop saturating beyond $\pm$2 T, with no remanence and zero coercive field.
Moreover, no differences have been detected between normal and grazing incidence (\textit{Supporting Information}, Figure \ref{fig:xmcd_Au_GI_NI}).
Low remanence and small coercive field are recurring characteristics in 2D vdW ferromagnets. In the specific, the coercive field diminishes with the layer number and tends to vanish when approaching the monolayer thickness \cite{li_CrSe2}.
It is worth noting that the coercive field measured by \textit{Bedoya-Pinto et al.} at 7.5 K and 8 T is of the order of 10 mT \cite{bedoya}. 
The intrinsic remanence of the beamline superconducting magnet and the TEY signal noise while sweeping magnetic field across the zero severely limit the sensitivity of the measurements in this range and might have prevented from reproducing the same results.
Despite that, a finite magnetic dichroism of $\sim$6\% persists in the XMCD spectra measured in remanence after the application of a 9 T magnetic field, indicating a net residual magnetic moment (\textit{Supporting Information}, Figure \ref{fig:xmcd_Au_0T}). 
In this respect, as grown SL CrCl$_3$/Au(111) can be described as a particularly \textit{soft} ferromagnet.
The repetition of the magnetization measurements at 4, 15 and 25 K demonstrates that the low-temperature S-shape profile evolves rapidly towards a linear dependence by increasing the temperature, suggesting that the onset of the FM order occurs within the 5-15 K temperature window (\textit{Supporting Information}, Figure \ref{fig:T_dep}).

Figure \ref{fig:Au}f reports the polarization-dependent XAS and the XMCD spectra calculated using LFM theory with the \textit{Quanty} code. 
Experimental conditions of beam incidence, temperature and external applied magnetic field, respectively set to NI, 4 K, and 9 T, have been recreated in the simulations. 
A $D_{3d}$ symmetry has been adopted, which addresses the trigonal distortion of the pristine octahedral ($O_{h}$) coordination within the CrCl$_6$ cluster unit. The simulations are sensitive mostly to the crystal field interaction, the hybridization potentials and the exchange field.
These and other details are reported in the \textit{Supporting Information}. 
The simulations allow to estimate some noteworthy expectation values that cannot be retrieved experimentally such as crystal field and magnetic moments. 
The calculated lineshapes optimally agree with the experimental results. To this purpose, octahedral and trigonal crystal field interactions have been optimized at 1.75 and 0.20 eV. 
The calculated spin angular momentum projection along the z-axis was found to be $\left< Sz \right> = 2.98 \mu_B$, in agreement with the nominal value for a high spin Cr $d^3$ configuration and the value extrapolated with corrected sum rules by \textit{Bedoya-Pinto et al.}  \cite{bedoya}. 
The orbital moment is instead nearly quenched ($\left< Lz \right> = -0.04 \mu_B$). 
These results confirm that the magnetic moment in CrCl$_3$ is purely spin-based  \cite{mcguire_CrCl3}.

%%%%%%%%%%%%%%%%%%%%%%%%%%%%%%%%%%%%%%%%%%%%%%%%%%%%%%%%%%%%%%%%%%%%%
%% on Ni111
%%%%%%%%%%%%%%%%%%%%%%%%%%%%%%%%%%%%%%%%%%%%%%%%%%%%%%%%%%%%%%%%%%%%%

\subsection{CrCl$_2$/Ni(111)}
\begin{figure}
    \centering
    \includegraphics[scale=0.8]{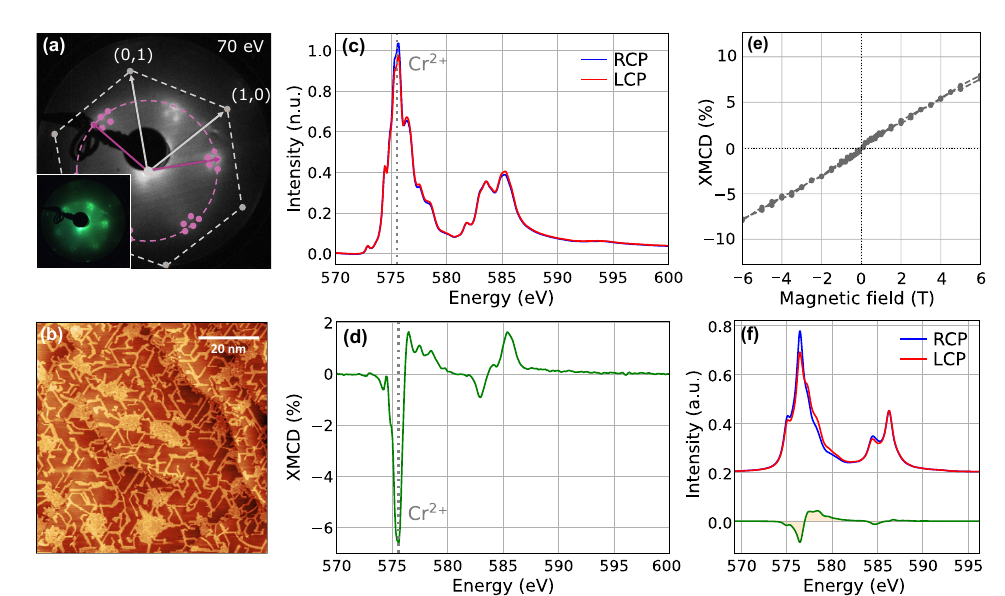}
    \caption{
    \textbf{CrCl$_{2}$/Ni(111)} \textbf{(a)} LEED pattern acquired at 70 eV. Brown spots and vectors correspond to substrate reflections and unit cell, while violet ones indicate the overlayer reconstruction; the dashed circumference highlight the ring signal crossing the overlayer spots. \textit{inset}: same LEED image without guidelines. \textbf{(b)} STM image (80$\times$80 nm$^2$) of CrCl$_2$ islands grown on the Ni(111) surface. \textbf{(c-d)} XAS and XMCD spectra taken around Cr L$_{3,2}$ edges at 4 K, under 9 T magnetic field applied along the beam direction, and in NI; the RCP and LCP XAS intensities are normalized to the average; the XMCD intensity is expressed in percentage units relative to the average XAS. \textbf{(e)} Magnetization curves measured with 575.5 eV photon energy (Cr$^{2+}$ edge, indicated by the dashed line in (c) and (d)), at 4 K and in NI. \textbf{(f)} Simulations of the experimental XAS and XMCD spectra shown in (c) and (d).
    }
    \label{fig:Ni}
\end{figure}

Ni(111) was chosen as a magnetic substrate for the growth of Cr Chloride monolayers to probe possible magnetic coupling effects. However, opposite to Au, Ni manifests a stronger chemical affinity towards ligand species such as chalcogens or halogens \cite{sant_WSe2}. The growth process, similar to the CrCl$_3$/Au(111) case, is described in the \textit{Experimental Section}.
The LEED pattern of the as grown sample is shown in Figure \ref{fig:Ni}a. It is characterized by six sets of spot quintuplets lying across a faint intensity ring internal to the hexagon marking the substrate reflections. 
The quintuplets are 30$^{\circ}$ rotated with respect to the substrate lattice vectors, and are interpreted as the overlap of six equivalent 2D crystal domains having a surface unit cell with lattice constants a=b=3.60 \AA and internal angle $\alpha$=50$^{\circ}$. This value is compatible with the one calculated for CrCl$_2$ \cite{bo}, however experimental reference data on 2D epitaxial compounds are still missing.
The LEED pattern has been simulated with the \textit{LEEDpat4} software \cite{leedpat} and the results are reported in Figure \ref{fig:LEED_Ni} of the \textit{Supporting Information}.

The STM measurements show that the overlayer grows in the form of interconnected ribbons with preferential directions at 120$^{\circ}$ that clusterize at the merging points (Figure \ref{fig:Ni}b). 
A coverage of about 50\% and an average thickness of $\sim$3 \AA were estimated. 
Stripe-shaped islands were observed also for CrI$_2$ on graphite and other dihalides epitaxially grown on Au(111) \cite{li_CrI3}.

The XAS and XMCD spectra relative to this sample are shown in Figures \ref{fig:Ni}c-d. The XAS whitelines are different from those measured on the Au(111) substrate (Figures \ref{fig:Au}c-d), and both L$_3$ and L$_2$ edges are shifted to lower energies by 1.5 eV, respectively at 575.5 eV and 584.0 eV. 
The new L$_3$ edge position can be associated to a lower Cr$^{2+}$ oxidation state. 
This result suggests that the Ni substrate has reduced the CrCl$_3$ clusters nucleated on the surface, enabling the formation of CrCl$_2$ species. 
Compared to CrCl$_3$/Au(111), the XMCD signal has much lower intensity ($\sim$6\% of the average XAS intensity) but similar lineshape. 

The magnetization curve (Figure \ref{fig:Ni}e), measured by tuning the probe photon energy to the dichroic peak at 575.5 eV, displays a purely linear dependence.
This trend is compatible with the behavior observed in AFM bulk CrI$_2$ in the same temperature window \cite{schneeloch}.

We simulated the polarization-dependent XAS and XMCD spectra of CrCl$_2$/Ni(111) for the same geometrical and environmental conditions described for CrCl$_3$/Au(111).
Fine agreement with the experiments was found in this case by adopting a pure octahedral symmetry (O$_h$) and the results are reported in Figure \ref{fig:Ni}f. 
Noteworthy, the best fit of the CrCl$_2$ XAS has been obtained for a lower $10Dq$ (1.50 eV) than the CrCl$_3$ case, and for zero exchange field, this latter information supporting the non-ferromagnetic behavior observed in the experiments. 
Also in this case, all the relevant parameters are reported in the \textit{Supporting Information}. 
Similarly to CrCl$_3$, the extrapolated orbital moment is quenched ($\left< Lz \right> = 0.004 \mu_B$), whereas the spin angular momentum projection along the z-axis was found to be $\left< Sz \right> = -0.275 \mu_B$, reflecting a low-spin case compatible with the weak magnetism of this sample.

%%%%%%%%%%%%%%%%%%%%%%%%%%%%%%%%%%%%%%%%%%%%%%%%%%%%%%%%%%%%%%%%%%%%%
%% on graphene
%%%%%%%%%%%%%%%%%%%%%%%%%%%%%%%%%%%%%%%%%%%%%%%%%%%%%%%%%%%%%%%%%%%%%

\subsection{CrCl$_{2(3)}$/SLG/Ni(111)}
To hinder the chemical reactivity of Ni(111) and enable the formation of SL CrCl$_3$ monolayers - showing intense XMCD and FM properties - also on top of a magnetic substrate, the Ni(111) surface was passivated with SLG. 
SLG was prepared by ethylene (C$_2$H$_4$) gas pyrolysis at 570$^{\circ}$C, as described in the \textit{Experimental Section}. 
This method has been demonstrated successful in forming high quality graphene with domain extension limited by the lateral size of the substrate terraces \cite{batzill}. When residual patches of Nickel Carbide (Ni$_2$C) - possible byproduct of the reaction - are detected by LEED, a second post-growth annealing at 475$^{\circ}$C in UHV is required for the reconversion into graphene \cite{lahiri_carbide} (see \textit{Supporting Information} and Figure \ref{fig:Gr_growth}).

\begin{figure}
    \centering
    \includegraphics[scale=0.8]{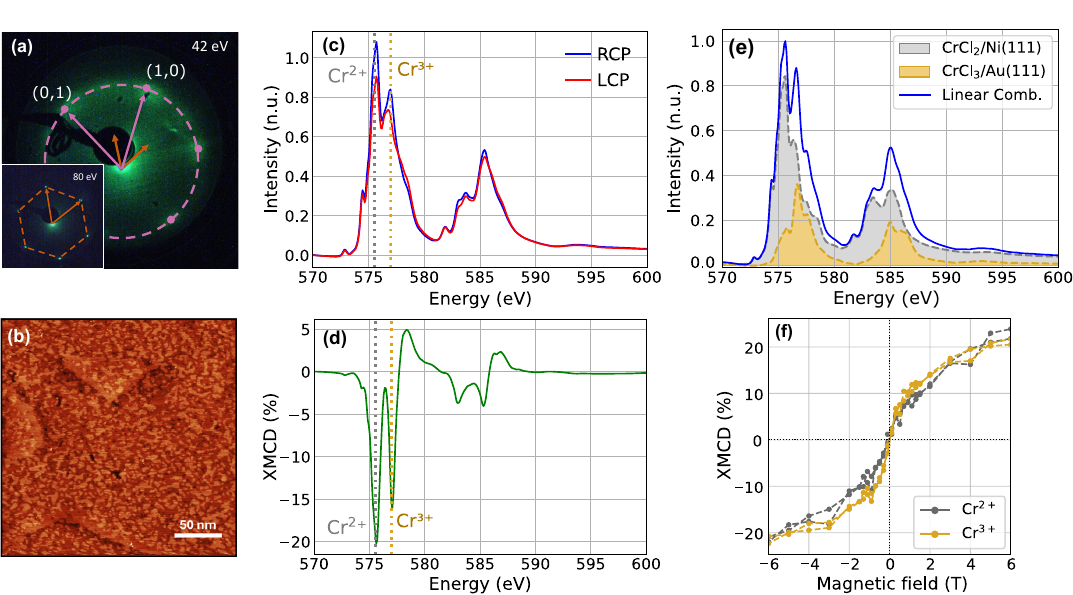}
    \caption{
    \textbf{CrCl$_{2(3)}$/SLG/Ni(111) (I).} \textbf{(a)} LEED pattern acquired at 42 eV. Violet spots and vectors correspond to overlayer reflections, while brown vectors indicate the substrate crystallographic axes directions (reflections are out of the image); the dashed circumference highlight the ring signal crossing the overlayer spots; \textit{inset}: LEED pattern of SLG/Ni(111) substrate acquired at 80 eV. \textbf{(b)} STM image (200$\times$200 nm$^2$) measured after CrCl$_{3}$ deposition on SLG/Ni(111). \textbf{(c-d)} XAS and XMCD spectra taken around Cr L$_{3,2}$ edges at 4 K, under 9 T magnetic field applied along the beam direction, and in NI; the RCP and LCP XAS intensities are normalized to the average; the XMCD intensity is expressed in percentage units relative to the average XAS. \textbf{(e)} Experimental polarization average XAS of CrCl$_3$/Au(111) (gold) and CrCl$_2$/Ni(111) (gray) samples and their linear combination (blue). \textbf{(f)} Magnetization curves taken at 4 K and in NI at the Cr$^{2+}$ and Cr$^{3+}$ edge energies indicated by the dashed lines in (c) and (d).
    }
    \label{fig:Gr_kurt}
\end{figure}

% Kurt's graphene
Cr chloride grows on SLG/Ni(111) in the form of small tens of nm large flakes of irregular shapes distributed all over the SLG surface (Figure \ref{fig:Gr_kurt}b). 
As consequence of the short range structural order and the morphological disorder, the LEED pattern shows a diffused background signal behind very broad and weak reflections oriented at 30$^{\circ}$ from the Ni(111) axes (Figure \ref{fig:Gr_kurt}a). 
The morphology of the chloride flakes as observed by STM looks different from the monolayers grown on Au(111) and Ni(111). 
The lateral size of the 2D Chlorides could have been limited by the low diffusion rate of the adsorbed nuclei or by the quality of the SLG layer underneath, \textit{e.g.} the presence of point defects or the high density of domain boundaries.

Likewise the previous samples, Figures \ref{fig:Gr_kurt}c-d report the XAS and XMCD spectra acquired at 4 K, 9 T and in NI for the Cr Chlorides grown on SLG/Ni(111). 
The XAS whiteline reproduces mostly the spectral shape acquired for CrCl$_2$/Ni(111) (Figure \ref{fig:Ni}c) with the main L$_{3}$ absorption edge and the relative XMCD peak sitting exactly on the Cr$^{2+}$ energy position, \textit{i.e.} 575.5 eV. 
Noteworthy, the XAS profile along the L$_3$ post-edge region appears broader than in Figure \ref{fig:Ni}c, while in the XMCD spectrum an additional peak emerges, which sits on the Cr$^{3+}$ edge energy, \textit{i.e.} at 577.0 eV. 
The double dichroism is likewise replicated on the L$_2$ peak, but was not present in the CrCl$_2$ spectrum measured on the bare Ni(111) substrate (Figure \ref{fig:Ni}d).
From this evidence, we deduce that both CrCl$_2$ and CrCl$_3$ phases coexist on the SLG/Ni(111) surface.

Focusing on the relative XMCD intensity of the two phases, it can be noted how CrCl$_3$, generates a magnetic response comparable to that of CrCl$_2$. The latter is however more abundant in the sample, as indicated by the overall Cr$^{2+}$ character of the XAS lineshape. 
A quantitative estimation of the two phases amount was obtained by fitting the experimental average XAS with the linear combination of the CrCl$_3$ and CrCl$_2$ XAS lineshapes shown in Figures \ref{fig:Au}c and \ref{fig:Ni}c, respectively measured on bare Au(111) and Ni(111), after polarization averaging.
The weight factors associated to the single contributions reproduce the ratio between the CrCl$_2$ and CrCl$_3$ species.
We estimated that the CrCl$_3$ species present on the sample is comprised between the 20 and 25\% of the total overlayer amount.
The calculated sum spectrum fitting best the experimental XAS together with the CrCl$_3$ and CrCl$_2$ components are shown in Figure \ref{fig:Gr_kurt}e.

The results described so far for the two Ni-based substrates (bare and passivated by SLG) are reproducible: the Cr$^{3+}$ peak is observed on Ni(111) only in presence of SLG, whereas it is not found when the growth is carried out on the bare surface. 
This fact supports the role of SLG in protecting the overlayer by physically decoupling it from the reactive Ni substrate \cite{sant_WSe2}.
CrCl$_3$ reduction to CrCl$_2$ does not occurs instead on the Au(111) surface, owing to the chemical inertness of the substrate. 

We then measured the magnetization loops on SLG/Ni(111) using two different photon probes tuned at the Cr$^{2+}$ and Cr$^{3+}$ absorption edges (more details are given in the \textit{Experimental Section}). 
\textit{Site-selective} measurements allowed us to address the single CrCl$_2$ and CrCl$_3$ phases separately and study the magnetic behavior of the different species found on this sample.  
The magnetization curves corresponding to the two energies are shown in Figure \ref{fig:Gr_kurt}f. They appear as two S-shaped closed loops with unsaturated magnetization even at high field ($\pm$6 T). These features are typically found in SPM nanostructures and originate from the nanometric size of the Chloride 2D particles, which act as independent magnetic entities dipped in the external magnetic field.
The loops are identical on both edges and the dichloride and trichloride species cannot be distinguished on the base of their magnetic response.

\begin{figure}
    \centering
    \includegraphics[scale=0.8]{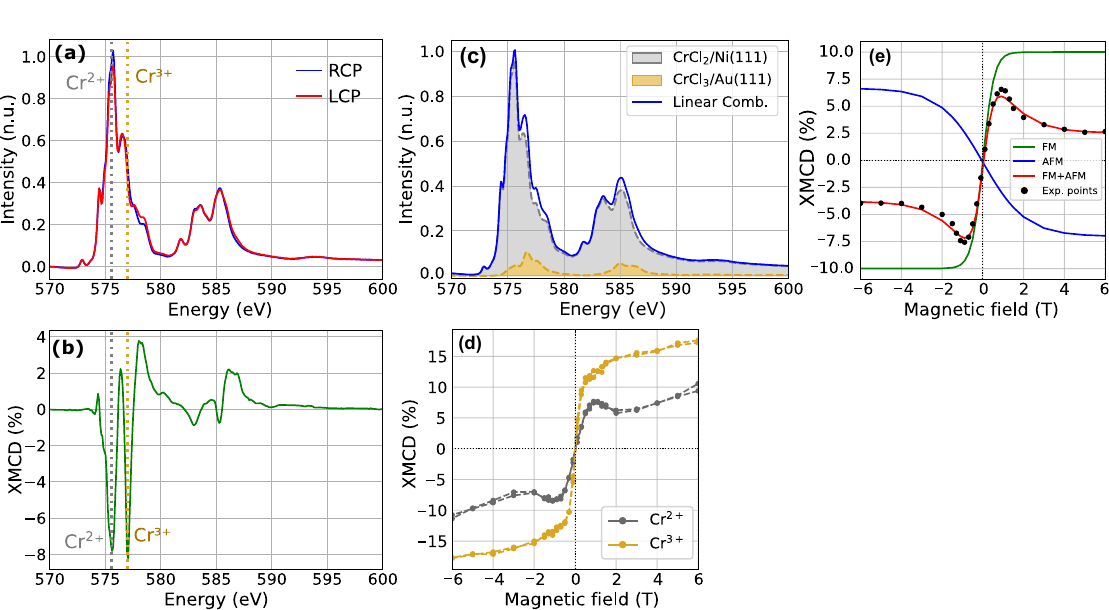}
    \caption{
    \textbf{CrCl$_{2(3)}$/SLG/Ni(111) (II).} \textbf{(a-b)} XAS and XMCD spectra taken around Cr L$_{3,2}$ edges at 4 K, under 9 T magnetic field applied along the beam direction, and in NI; the RCP and LCP XAS intensities are normalized to the average; the XMCD intensity is expressed in percentage units relative to the average XAS. \textbf{(c)} Experimental polarization average XAS of CrCl$_3$/Au(111) (gold) and CrCl$_2$/Ni(111) (gray) samples and their linear combination (blue). \textbf{(d)} Magnetization curves taken at 4 K and in NI at the Cr$^{3+}$ and Cr$^{2+}$ edge energies indicated by the dashed lines in (a) and (b). \textbf{(e)} Experimental points and fitting (in red) of the Cr$^{2+}$ magnetization loop as reported in (d) after linear background subtraction. The green and blue lines represent the \enquote{direct} and \enquote{reverse} magnetization fit components (see \textit{Experimental Section)}. 
    }
    \label{fig:Gr_cc}
\end{figure}

% graphene from carbide conversion
A second sample has been prepared instead on a SLG/Ni(111) obtained by conversion of the metastable Ni$_2$C formed on the Ni(111) surface during cooling from the pyrolysis temperature (see \textit{Experimental Methods}). 
The spectroscopic and magnetic measurements are reported in Figure \ref{fig:Gr_cc}.
The XAS measured around the Cr L$_3$ edge (Figure \ref{fig:Gr_cc}a) still presents an overall Cr$^{2+}$ character. However, the post-edge region appears less broad than the previous case (Figure \ref{fig:Gr_kurt}c).
This observation was quantitatively assessed by fitting the average XAS profile according to the method described above. We estimated a lower ($\sim$10\%) concentration of CrCl$_3$ phase in this sample (Figure \ref{fig:Gr_cc}c).
Nevertheless, the XMCD shape confirms the coexistence of CrCl$_2$ and CrCl$_3$ species also in this case (Figure \ref{fig:Gr_cc}b), as proved by the double dichroism at the Cr$^{2+}$ and Cr$^{3+}$ characteristic energies.

Remarkably, Figure \ref{fig:Gr_cc}d reveals that the magnetization curves acquired with the two different photon energies appear different, though conserving the closed loop character. At the Cr$^{3+}$ edge the magnetization shows a higher degree of squareness, similarly to the one measured for CrCl$_3$/Au(111). 
By tuning the photon energy to the Cr$^{2+}$ edge, the data show instead a complex shape, characterized by a prominent linear trend visible beyond $\pm$2 T, a steep field dependence across zero and two anomalous \textit{humps} at $\sim\pm1$ T.
This lineshape has been decomposed in Figure \ref{fig:Gr_cc}e in two opposite-in-sign S-shaped contributions (in green and blue colors), after the subtraction of a linear background for better visualization.
The two components present different saturation points and slopes across the zero field region. 
They have been modeled by two sigmoid functions (see \textit{Experimental Section}) and then linearly combined to fit the experimental data (red curve). 
In analogy to a nanoparticle or core-shell picture \cite{de}, we suggest that the \textit{direct} (green) or \textit{reverse} (blue) magnetization contributions (\enquote{direct} and \enquote{reverse} are referred to the Cr$^{3+}$ magnetization, as reported in Figure \ref{fig:Gr_cc}d) originate from spin-disorder associated to frustrated magnetic coupling between CrCl$_2$ domains surrounding the CrCl$_3$ ones. This interpretation is addressed in the next section.

%%%%%%%%%%%%%%%%%%%%%%%%%%%%%%%%%%%%%%%%%%%%%%%%%%%%%%%%%%%%%%%%%%%%%
%% Discussion
%%%%%%%%%%%%%%%%%%%%%%%%%%%%%%%%%%%%%%%%%%%%%%%%%%%%%%%%%%%%%%%%%%%%%

\subsection{Discussion}
% magnetic behaviour: SPM and frustration
Within a nanoparticle description justified by the nanometric size of the chloride domains as observed by STM, S-shape profile, zero coercive field and unsaturated magnetization, all point to a SPM behaviour of the chloride islands grown on SLG/Ni(111). 
In this picture, 2D nano-islands can be treated as single independent domain entities with a macroscopic \enquote{\textit{super-spin}} that aligns with the external magnetic field \cite{de,peddis}. This is the case observed on the first SLG/Ni(111) sample.
When the density of the domains increases, interaction between domains becomes more effective and collective behaviors associated to competing order manifest.
Frustration is not unusual in mixed Cr Chloride/graphene systems, and was already reported for CrCl$_3$ intercalated graphite, as well as in CrCl$_3$ chains confined within carbon nanotubes, which may be considered respectively as the 3D and 1D counterparts of the 2D system under study \cite{suzuki,li_swcnt}.
In those systems, Cr$^{2+}$ species are generated by the charge transfer from the graphite sheets towards the CrCl$_3$ islands, eventually reducing them. Such hypothesis can be reasonably extended to our case study.
In the second SLG/Ni(111) sample, graphene sheets do not form directly, but following the annealing necessary to convert the metastable carbide phase. We believe that this process leads to defective graphene, mainly characterized by heterogeneous and fragmented domains and a high density of domain boundaries. These narrow domains may have affected the final morphology of the chloride overlayer in a similar way, determining the formation of small magnetic \enquote{\textit{spin clusters}}, acting as weakly interacting SPM units \cite{peddis,de}.
In the ensemble of closely packed 2D spin clusters, random distances and anisotropic dipolar interactions lead to the competition among FM and AFM interactions, which gives rise to the parallel and antiparallel alignment (\textit{i.e.} the \enquote{direct} and \enquote{reverse} magnetization loops) between SPM CrCl$_2$ and CrCl$_3$ domains \cite{ma}.

% Cr2+ stabilization on Gr
Previously, we have explicitly formulated the hypothesis that CrCl$_2$ formation on SLG/Ni(111) occurs due to the electron charge transferred from the graphene layer, similarly to other cases reported in the literature \cite{suzuki,li_swcnt}.
In fact, despite the fact that the graphene passivation leads to a significant amount of CrCl$_3$ unachiavable on bare Ni(111), CrCl$_2$ still represents the dominant phase found on the sample.
Besides this hypothesis, other possible mechanisms for CrCl$_2$ formation were considered. 
Firstly, the Cr reduction by Ni atoms could be in this circumstance only a partial explanation, since it requires the direct contact between the two materials, whereas we expect a relatively small portion of uncovered areas on the Ni(111) surface after SLG growth.
Second, past works suggest that the surface thermodynamics of halide species deposited on SLG/metal implies a variety of possible microscopic events, including trihalide dissociation into dihalide compounds and halogen atoms \cite{generalov,vinogradov}. The latter might finally intercalate SLG. 
All these processes are thermally activated and they can be controlled by means of the annealing temperature, which is typically reported between 150 and 200$^{\circ}$C.  \cite{generalov,vinogradov} 
However, during the chloride growths, the substrate was maintained at temperature lower than 150$^{\circ}$C to prevent CrCl$_3$ dissociation.
Actually, the mechanism enabling CrCl$_3$ dissociation and SLG intercalation by Cl and Cr atoms cannot be fully established with our techniques.
However, our STM images do not show profile modulation and roughness such as due to localized Cl nanodomains intercalated between the SLG and the metallic substrate \cite{sant_MoS2}. 
Moreover, reconstructions of intercalated adatoms do not appear in the LEED pattern upon growth, though they could be invisible due to structural disorder.
Charge transfer from SLG towards Chlorides remains therefore a plausible argument to explain the coexistence of CrCl$_2$ and CrCl$_3$ on top of SLG. 
In this respect, it is reasonable to assume that CrCl$_2$ formation occurs randomly within the CrCl$_3$ layers, \textit{i.e.} there is no net CrCl$_2$ and CrCl$_3$ phase segregation, but a \enquote{puzzle} of patches separated by a dense network of grain boundaries.
The simultaneous formation of MX$_2$ and MX$_3$ species on the same substrate at $\sim$ 100$^{\circ}$C is possible by MBE growths due to the narrow stability temperature windows of the two species \cite{li_CrI3,li_lateral}.

% the proximity effect of the Ni substrate
Finally, we address the role of magnetic Ni(111) substrate in the as built vdW stacking. 
One may wonder in fact whether the magnetism in the overlayer could be controlled through the heterostructure stacking, \textit{i.e.} by the modulation via SLG of the interlayer exchange between the Ni(111) substrate and the 2D Chlorides.
For instance, in 0D magnetic molecular systems deposited on SLG/Ni(111), the spin alignment of the metal center is determined by the coupling distance.
Epitaxial SLG is used to reduce the out-of-plane exchange interaction by increasing the separation between substrate and molecules, and transform the FM exchange into AFM \cite{corradini}.
However, in the hysteresis cycles measured on such systems, the saturation point is the same for the overlayer and the substrate and their XMCD sign is parallel or antiparallel as a function of their mutual distance.  
In our case instead, no distance-modulated dichroism sign has been experimentally observed by XMCD between Ni and 2D Chloride, and no correlation between saturation fields in Cr and Ni magnetization has been detected (Ni has much smaller coercive field than those measured for Chlorides in this work).
We conclude that the Ni(111) substrate does not represent a coupling term for the grown SPM Chlorides, and SLG does not work as a suitable filter through which the magnetic properties of Cr Chloride can be controlled.

%%%%%%%%%%%%%%%%%%%%%%%%%%%%%%%%%%%%%%%%%%%%%%%%%%%%%%%%%%%%%%%%%%%%%
%% Conclusions
%%%%%%%%%%%%%%%%%%%%%%%%%%%%%%%%%%%%%%%%%%%%%%%%%%%%%%%%%%%%%%%%%%%%%

\section{Conclusion}
In this work we described the structural, morphological and magnetic nature of Cr Chloride (sub)monolayers grown by evaporation in UHV of the same CrCl$_3$ powder precursor on Au(111), Ni(111) and SLG/Ni(111) substrates. 
The as grown 2D materials were characterized by LEED, STM and XMCD using synchrotron soft X-rays. 
We showed that the two metallic substrates, \textit{i.e.} Au(111) and Ni(111), promote the formation of different Chloride species, \textit{i.e.} CrCl$_3$ and CrCl$_2$, showing respectively weak FM and non-FM behaviors. 
In order to hinder the chemical reactivity of the Ni substrate towards the CrCl$_3$ clusters upon their adsorption on the surface, the Ni(111) was passivated with SLG. 
We showed that SLG favors the partial formation of nanometric CrCl$_3$ islands in addition to CrCl$_2$ ones. The latter are probably generated by the electron transfer channel from SLG. 
The CrCl$_3$ amount is correlated to the SLG synthesis process and final morphology, and the highest estimated CrCl$_3$ concentration was found $\sim$25\%. 
The coexistence of CrCl$_2$ and CrCl$_3$ phases is evidenced by two dichroic peaks in the XMCD spectra matching the characteristic Cr$^{3+}$ and Cr$^{2+}$ energy positions, and by site-selective magnetization measurements performed by tuning the probing photons to the two absorption edges. 
The latter reveals the SPM nature of both CrCl$_2$ and CrCl$_3$ nano-domains, which is intimately related to the surface morphology of the SLG underneath. 
Moreover, the complex shape of the CrCl$_2$ magnetization acquired in one of the samples was explained as the sum of direct and reverse loops interpreted in terms of competing FM and AFM interactions between adjacent CrCl$_2$ and CrCl$_3$ SPM domains.
No evidence of SLG-mediated coupling between the Ni(111) substrate and any Chloride species was instead observed.
These results demonstrate the variety of possible chemical and magnetic phases that can be found on a single crystal substrate surface upon Chloride MBE growths, which adds to the already known complexity of the morphology at the microscopy level (moirée, domain shapes, presence of single and double layers, etc.).
The results invoke deeper and more specific microscopy studies to understand the mechanisms behind the phase formations, in view of more uniform structural and magnetic order in Chlorides and, more generally, in 2D epitaxial transition metal halides.

%%%%%%%%%%%%%%%%%%%%%%%%%%%%%%%%%%%%%%%%%%%%%%%%%%%%%%%%%%%%%%%%%%%%%
%% Methods
%%%%%%%%%%%%%%%%%%%%%%%%%%%%%%%%%%%%%%%%%%%%%%%%%%%%%%%%%%%%%%%%%%%%%

\section{Experimental Section}

\threesubsection{Sample growth}
Sample growth and preliminary \textit{in situ} characterizations were carried out in the MBE setup of the XMCD endstation at the ESRF-ID32 beamline \cite{brookes,kummer}. 
The metallic Au(111) and Ni(111) substrates were cleaned using standard sputtering (1 kV, $1\times10^{-5}$ mbar of Ar pressure) and annealing (550-600$^{\circ}$C) cycles.

SLG was grown on Ni(111) by ethylene (C$_2$H$_4$) pyrolysis. During the process, the substrate temperature was kept at around 570$^{\circ}$C and the ethylene gas exposure was estimated between 500-650 L (Langmuir). 
The next steps were carried out by following two different variants, (\textit{i.e.}) by operating the cooling ramp in vacuum or under ethylene atmosphere. 
In the first case, residual carbide patches on the surface require a post-annealing treatment (at 475$^{\circ}$C) for conversion into SLG, as described by \textit{Lahiri et al.} \cite{lahiri_carbide}. This method led to a more defective graphene, with heterogeneous coverage. 
In the second case, \textit{i.e.} by slowly cooling the sample in ethylene atmosphere, a more homogeneous SLG is obtained \cite{kummer}. 

The growth of CrCl$_3$ monolayers was conducted in UHV environment (background pressure below 10$^{-9}$ mbar) by MBE using a Knudsen cell evaporator filled with anhydrous CrCl$_3$ powder (provided by Sigma Aldrich, 99.99\% pure) heated up to 530$^{\circ}$C. 
The deposition rate was calibrated \textit{a posteriori} by STM to 0.10 ML/min, and the time adjusted for the desired target coverage. 
Both on Au(111) and Ni(111), a minimum sub-monolayer coverage of 50$\%$ was achieved. 
The substrate temperature was kept below 150$^{\circ}$C to limit the substrate reactivity towards the CrCl$_3$ precursor species and to avoid halogen desorption.

Deposition and substrate contamination were regularly checked by Auger electron spectroscopy (AES) after each growth or sputtering/annealing cycle.
All the samples were characterized \textit{in situ} by LEED and STM prior to the X-ray measurements.

\vskip 0.1in

\threesubsection{X-ray measurements}
XAS/XMCD experiments have been carried out at the XMCD station of the ID32 beamline at the ESRF \cite{brookes,kummer}. 
The measurements were performed inside the High Field Magnet (HFM) chamber, in a background pressure lower than 3$\times$10$^{-10}$ mbar, by scanning the energy around the Cr L$_{3,2}$ edges using right- and left-circularly polarized (RCP, LCP) X-rays. 
The synchrotron beam presents a degree of polarization of almost 100$\%$ and the monochromator has a resolving power better than 5000. 
The samples were studied in the 4-300 K temperature with magnetic field varying between 0 and 9 T, applied along the beam direction. Measurements were performed both in normal (90$^{\circ}$) and grazing (30$^{\circ}$) incidence geometries.  
The absorption signal was measured in total-electron yield (TEY) mode and normalized by the intensity collected on a gold mesh placed in front of the sample stage.
Polarization-dependent XAS are plotted after normalization to the average XAS intensity.
XMCD intensity is plotted in percentage units relative to the average XAS intensity.
\textit{Site selective} magnetization curves were obtained by measuring the XMCD spectra at field variance between 6 T and -6 T, and between -6 T and 6 T, and then by integrating the signal under the selected Cr$^{2+}$ or Cr$^{3+}$ peaks (integration width: 0.25 eV).
The exact magnetization is proportional to the plotted XMCD signal.
The sigmoid function used to fit the magnetization curve corresponding to the Cr$^{2+}$ edge in Figure \ref{fig:Gr_cc}d-e is:  

\[ f(a, b, k_1, k_2) = \frac{a}{1 + e^{k_{1}x}} + \frac{b}{1 + e^{k_{2}x}} \]

where $a$, $b$, $k_1$, and $k_2$ model in order the saturation magnetization of each curve and their steepness.

%%%%%%%%%%%%%%%%%%%%%%%%%%%%%%%%%%%%%%%%%%%%%%%%%%%%%%%%%%%%%%%%%%%%%
%% Acknowledgements
%%%%%%%%%%%%%%%%%%%%%%%%%%%%%%%%%%%%%%%%%%%%%%%%%%%%%%%%%%%%%%%%%%%%%

\section{Acknowledgements}
We acknowledge the European Synchrotron Radiation Facility (ESRF) for provision of synchrotron radiation facilities under proposal number HC-5442 and we would like to thank all the past and present ID32 beamline staff for the technical and scientific support. This work has been partially performed within the MUSA-Multilayered Urban Sustainability Action project, funded by the European Union-NextGenerationEU, under the National Recovery and Resilience Plan (NRRP) Mission 4 Component 2 Investment Line 1.5: Strenghtening of research structures and creation of R\&D \enquote{innovation ecosystems}, set up of \enquote{territorial leaders} in R\&D. The authors thank in particular Prof. A. Brambilla from Politecnico di Milano and Dr. K Kummer from ESRF-ID32 for the fruitful discussions.

%%%%%%%%%%%%%%%%%%%%%%%%%%%%%%%%%%%%%%%%%%%%%%%%%%%%%%%%%%%%%%%%%%%%%
%% References
%%%%%%%%%%%%%%%%%%%%%%%%%%%%%%%%%%%%%%%%%%%%%%%%%%%%%%%%%%%%%%%%%%%%%

\medskip

\bibliographystyle{MSP}
\bibliography{main}
\newpage

%%%%%%%%%%%%%%%%%%%%%%%%%%%%%%%%%%%%%%%%%%%%%%%%%%%%%%%%%%%%%%%%%%%%%
%% Figures
%%%%%%%%%%%%%%%%%%%%%%%%%%%%%%%%%%%%%%%%%%%%%%%%%%%%%%%%%%%%%%%%%%%%%

\renewcommand{\figurename}{\textbf{Figure}}

\clearpage

\clearpage

\clearpage

\clearpage

%%%%%%%%%%%%%%%%%%%%%%%%%%%%%%%%%%%%%%%%%%%%%%%%%%%%%%%%%%%%%%%%%%%%%
%% Table of Contents
%%%%%%%%%%%%%%%%%%%%%%%%%%%%%%%%%%%%%%%%%%%%%%%%%%%%%%%%%%%%%%%%%%%%%

\begin{figure}
\textbf{Table of Contents}\\
  \medskip
  \centering
  \includegraphics[scale=1.0]{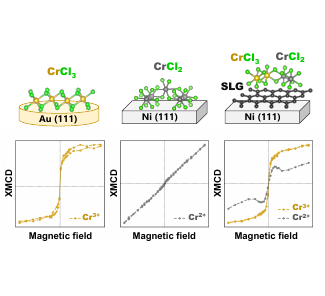}
  \medskip
  \caption*{
  Monolayer CrCl$_3$ and CrCl$_2$ can be grown by MBE respectively on Au(111) and Ni(111) substrates by evaporation of the same CrCl$_3$ precursor. 
  A combination of in situ techniques show that these materials present different structure (LEED), morphology (STM) and magnetic properties (XMCD). 
  On graphene-passivated Ni(111) the formation of both species is allowed, whose properties are governed by short-range order and magnetic frustration.
  }
\end{figure}

%%%%%%%%%%%%%%%%%%%%%%%%%%%%%%%%%%%%%%%%%%%%%%%%%%%%%%%%%%%%%%%%%%%%%
%% Supplemetary Figures
%%%%%%%%%%%%%%%%%%%%%%%%%%%%%%%%%%%%%%%%%%%%%%%%%%%%%%%%%%%%%%%%%%%%%

\begin{appendices}

\renewcommand\thefigure{\thesection S\arabic{figure}}
\setcounter{figure}{0}

\begin{figure}
    \captionsetup{labelformat=empty}
    \caption{}
    \label{fig:xmcd_bulk}
\end{figure}

\begin{figure}
    \captionsetup{labelformat=empty}
    \caption{}
    \label{fig:LEED_Au}
\end{figure}

\begin{figure}
    \captionsetup{labelformat=empty}
    \caption{}
    \label{fig:LEED_Ni}
\end{figure}

\begin{figure}
    \captionsetup{labelformat=empty}
    \caption{}
    \label{fig:z_profile}
\end{figure}

\begin{figure}
    \captionsetup{labelformat=empty}
    \caption{}
    \label{fig:Gr_growth}
\end{figure}

\begin{figure}
    \captionsetup{labelformat=empty}
    \caption{}
    \label{fig:xmcd_Au_0T}
\end{figure}

\begin{figure}
    \captionsetup{labelformat=empty}
    \caption{}
    \label{fig:xmcd_Au_GI_NI}
\end{figure}

\begin{figure}
    \captionsetup{labelformat=empty}
    \caption{}
    \label{fig:T_dep}
\end{figure}

\begin{figure}
    \captionsetup{labelformat=empty}
    \caption{}
    \label{fig:sim_CrCl2}
\end{figure}

\begin{figure}
    \captionsetup{labelformat=empty}
    \caption{}
    \label{fig:sim_CrCl3}
\end{figure}
    
\end{appendices}

%%%%%%%%%%%%%%%%%%%%%%%%%%%%%%%%%%%%%%%%%%%%%%%%%%%%%%%%%%%%%%%%%%%%%
%% End
%%%%%%%%%%%%%%%%%%%%%%%%%%%%%%%%%%%%%%%%%%%%%%%%%%%%%%%%%%%%%%%%%%%%%

\end{document}

\pagestyle{plain}
\rhead{\includegraphics[width=2.5cm]{vch-logo.png}}

%%%%%%%%%%%%%%%%%%%%%%%%%%%%%%%%%%%%%%%%%%%%%%%%%%%%%%%%%%%%%%%%%%%%%
%% Title
%%%%%%%%%%%%%%%%%%%%%%%%%%%%%%%%%%%%%%%%%%%%%%%%%%%%%%%%%%%%%%%%%%%%%

\title{In situ growth and magnetic characterization of Cr Chloride monolayers}
\maketitle
\vskip 0.2in
\title{Supporting Information}
\maketitle
\vskip 0.8in

%%%%%%%%%%%%%%%%%%%%%%%%%%%%%%%%%%%%%%%%%%%%%%%%%%%%%%%%%%%%%%%%%%%%%
%% Authors and affiliations
%%%%%%%%%%%%%%%%%%%%%%%%%%%%%%%%%%%%%%%%%%%%%%%%%%%%%%%%%%%%%%%%%%%%%

\author{Giuseppe Buccoliero,} 
\author{Pamella Vasconcelos Borges Pinho,}
\author{Marli dos Reis Cantarino,}
\author{Francesco Rosa,}
\author{Nicholas B. Brookes,}
\author{Roberto Sant}*

\begin{affiliations}
\vskip 0.2in
G. Buccoliero, Dr. P.V.B. Pinho, Dr. M.R. Cantarino, Dr. N.B. Brookes, Dr. R. Sant\\
ESRF, The European Synchrotron, 71 Avenue des Martyrs, CS40220, 38043 Grenoble Cedex 9, France
\vskip 0.1in
Dr. P.V.B. Pinho\\\
Université Paris-Saclay, CEA, Service de Recherche en Corrosion et Comportement des Matériaux, F-91191 Gif-sur-Yvette, France
\vskip 0.1in
Dr. R. Sant, F. Rosa\\
Dipartimento di Fisica, Politecnico di Milano, Piazza Leonardo da Vinci 32, I-20133 Milano, Italy.\\
Email Address: roberto.sant@polimi.it
\end{affiliations}

\clearpage

%%%%%%%%%%%%%%%%%%%%%%%%%%%%%%%%%%%%%%%%%%%%%%%%%%%%%%%%%%%%%%%%%%%%%
%% Bulk CrCl3
%%%%%%%%%%%%%%%%%%%%%%%%%%%%%%%%%%%%%%%%%%%%%%%%%%%%%%%%%%%%%%%%%%%%%
    
\section{XAS and XMCD of bulk CrCl$_3$}

\begin{figure}[h]
    \centering
    \includegraphics[scale=0.9]{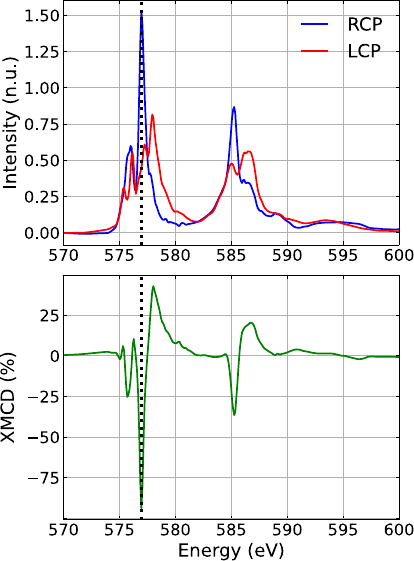}
    \renewcommand{\figurename}{\textbf{Figure}}
    \renewcommand{\thefigure}{\textbf{S\arabic{figure}}}
    \caption{
    XAS and XMCD spectra of a bulk CrCl$_3$ sample taken around Cr L$_{3,2}$ edges at 30 K, 9 T magnetic field applied along the beam and in NI.
    }
    \label{fig:xmcd_bulk}
\end{figure}

A commercially available \cite{hqgraphene} bulk CrCl$_3$ crystal was measured as reference for comparison against the epitaxially grown monolayer samples. 
The measurements reported in Figure \ref{fig:xmcd_bulk} have been performed at 30 K, 9 T and in normal incidence (NI). CrCl$_3$ is an in-plane antiferromagnetic van der Waals (vdW) material characterized by antiparallel alignment of spin momenta in adjacent planes. The single layer is instead intrinsically ferromagnetic  \cite{mcguire_CrCl3}. 
The XAS spectra taken with right- and left-circular polarized (RCP, LCP) X-rays, as well as the XMCD shape and intensity, closely reproduce those acquired for monolayer CrCl$_3$/Au(111) in Figure 1c-d.

%%%%%%%%%%%%%%%%%%%%%%%%%%%%%%%%%%%%%%%%%%%%%%%%%%%%%%%%%%%%%%%%%%%%%
%% Complementary information about in situ characterization of CrCl$_3$/Au(111) and CrCl$_2$/Ni(111)
%%%%%%%%%%%%%%%%%%%%%%%%%%%%%%%%%%%%%%%%%%%%%%%%%%%%%%%%%%%%%%%%%%%%%

\section{Complementary information about in situ LEED and STM characterizations of CrCl$_3$/Au(111) and CrCl$_2$/Ni(111)}

\begin{figure}[ht]
    \centering
    \includegraphics[scale=0.10]{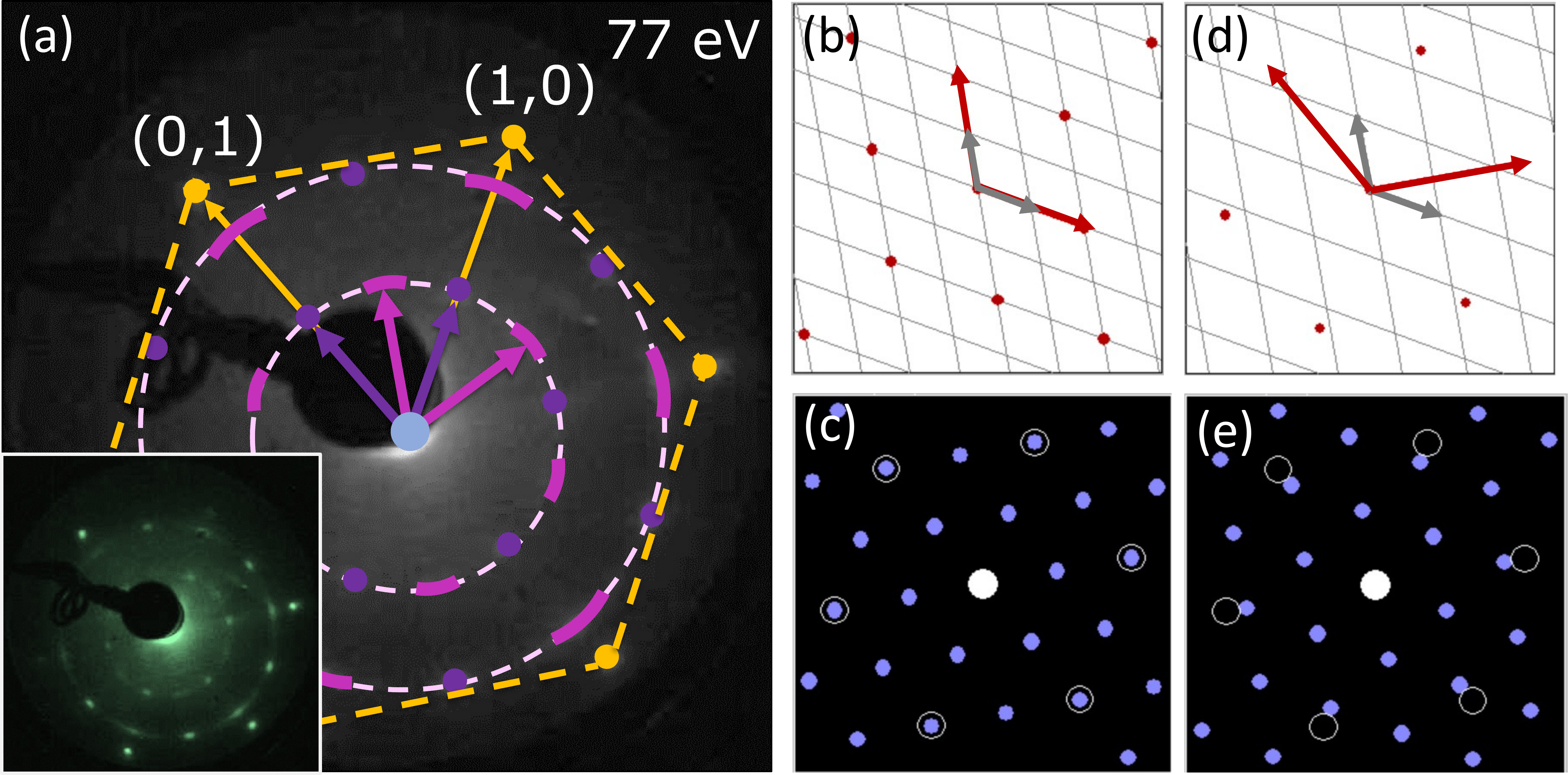}
    \renewcommand{\figurename}{\textbf{Figure}}
    \renewcommand{\thefigure}{\textbf{S\arabic{figure}}}
    \caption{\textbf{(a)} LEED pattern of CrCl$_3$/Au(111) acquired at 77 eV with guidelines. Yellow spots and vectors correspond to substrate reflections and unit cell, while pink and violet ones indicate the (2$\times$2) and (2$\times$2)R30$^{\circ}$ CrCl$_3$/Au(111) reconstructions; the white dashed circumferences highlight the ring signals crossing the overlayer spots; inset: as measured LEED pattern image without guidelines. \textbf{(b-c)} Real space lattice and reciprocal space simulations of the (2$\times$2)-CrCl$_3$/Au(111) reconstruction.  \textbf{(d-e)} Real space lattice and reciprocal space simulations of the (2$\times$2)R30$^{\circ}$-CrCl$_3$/Au(111) reconstruction.
    }
    \label{fig:LEED_Au}
\end{figure}

\begin{figure}[h]
    \centering
    \includegraphics[scale=0.10]{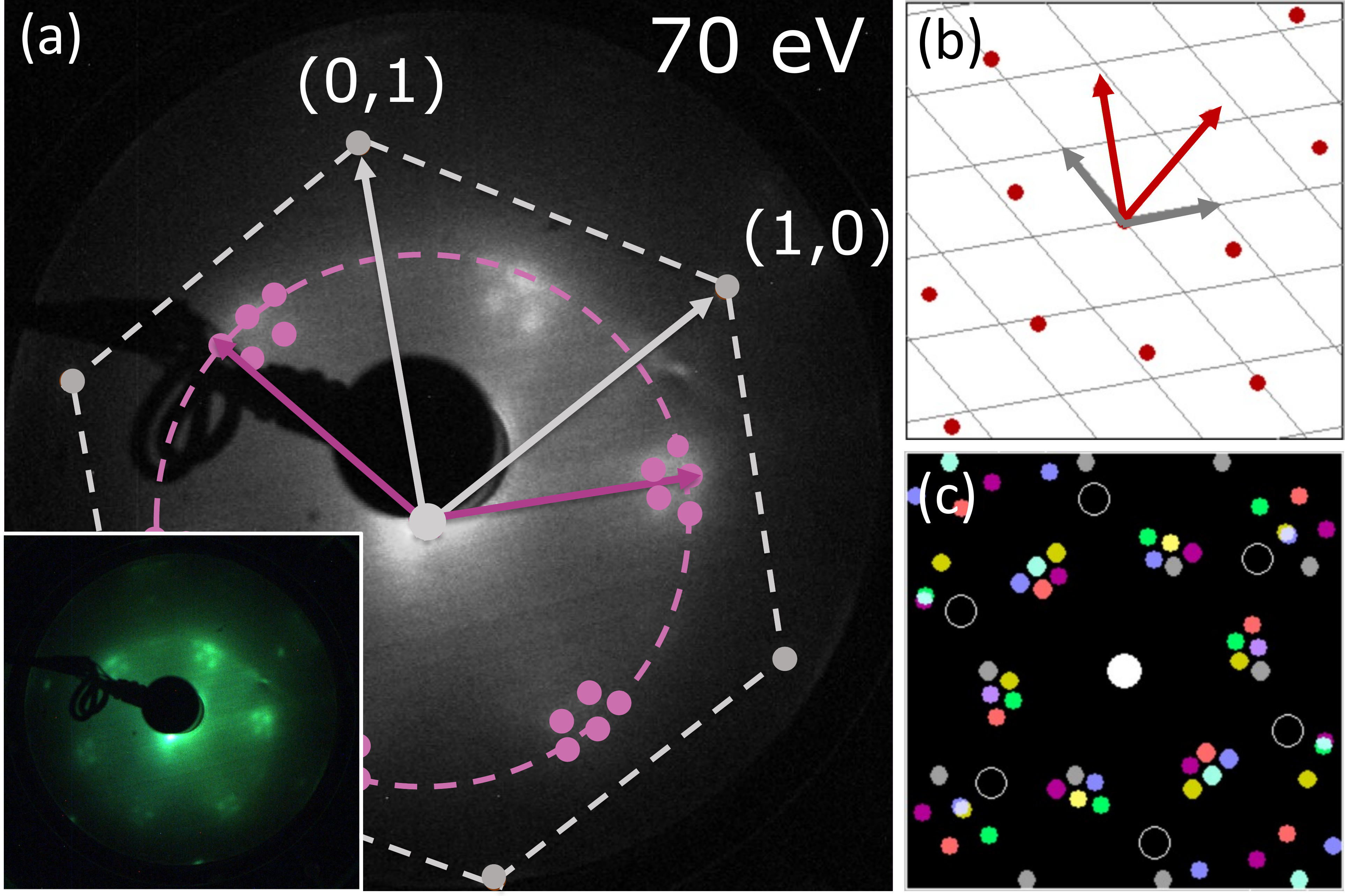}
    \renewcommand{\figurename}{\textbf{Figure}}
    \renewcommand{\thefigure}{\textbf{S\arabic{figure}}}
    \caption{\textbf{(a)} LEED pattern of CrCl$_2$/Ni(111) acquired at 70 eV with guidelines. Brown spots and vectors correspond to substrate reflections and unit cell, while violet ones indicate the overlayer reconstruction; the dashed circumference highlights the ring signal crossing the overlayer spots; inset: as measured LEED pattern image without guidelines; \textbf{(b-c)} Real space lattice and reciprocal space simulation of the CrCl$_2$/Ni(111) reconstruction.
    }
    \label{fig:LEED_Ni}
\end{figure}

The LEED patterns shown in Figures 1a and 2a in the main text are reproduced in Figure \ref{fig:LEED_Au}a and Figure \ref{fig:LEED_Ni}a together with real space and reciprocal space lattices simulated with the \textit{LeedPat4} software  \cite{leedpat}. 
The CrCl$_3$/Au(111) LEED pattern in Figure \ref{fig:LEED_Au}a is interpreted as the superposition of commensurate (2$\times$2) (Figure \ref{fig:LEED_Au}b-c) and incommensurate (2$\times$2)R30$^{\circ}$ (Figure \ref{fig:LEED_Au}d-e) reconstructions, the second characterized by larger mosaic spread.
The tabulated in-plane lattice parameters of Au(111) (2.88 \AA) and CrCl$_3$ (5.96 \AA) \cite{mcguire_CrCl3} are in fact close to a 1:2 ratio. We estimated a lattice constant of 5.8 \AA \ for the as grown CrCl$_3$/Au(111) sample.

The CrCl$_2$/Ni(111) LEED pattern is instead more complex. It is characterized by six sets of spot quintuplets lying across a faint intensity ring internal to the hexagon marking the substrate reflections. 
The quintuplets are 30$^{\circ}$ rotated with respect to the substrate lattice vectors.
They can be reproduced by adopting a surface cell with lattice constants $a=b=3.6$ \AA \ and internal angle $\alpha$=50$^{\circ}$ (Figure \ref{fig:LEED_Ni}b-c). The corresponding overlayer matrix is [[2.32,1.16],[-1.16,1.16]].
For both LEED images, the rings passing through the overlayer spots indicate a poor orientational alignment of the as-grown domains along the substrate crystallographic axes and point to a weak interface bonding typical of the vdW epitaxy.

\begin{figure}[h]
    \centering
    \includegraphics[scale=0.15]{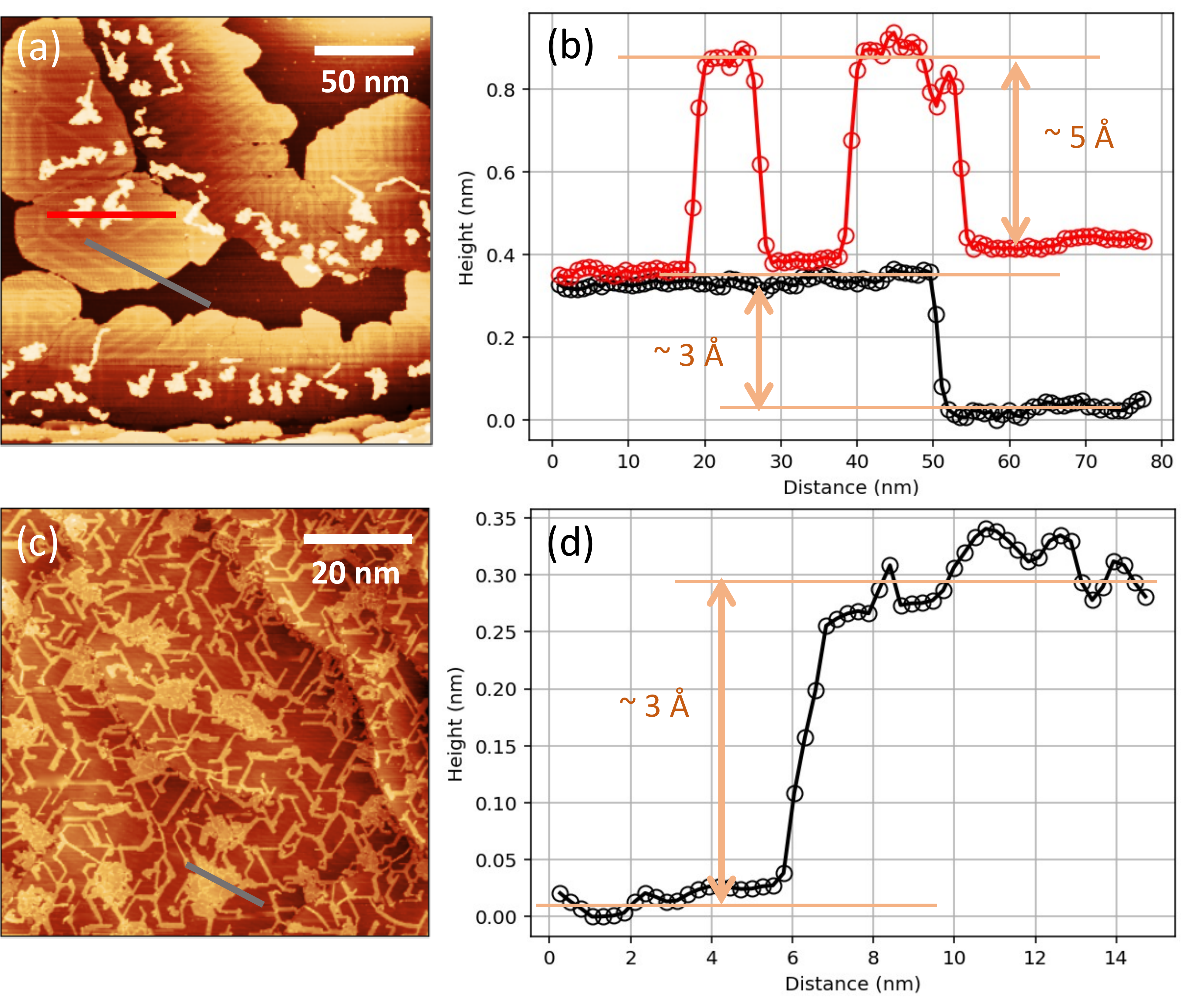}
    \renewcommand{\figurename}{\textbf{Figure}}
    \renewcommand{\thefigure}{\textbf{S\arabic{figure}}}
    \caption{\textbf{(a)} STM image (225$\times$225 nm$^2$) showing CrCl$_3$ on Au(111) surface. \textbf{(b)} Height profils along the directions marked with black and red in (a). \textbf{(c)} STM image (80$\times$80) showing CrCl$_2$ islands on Ni(111) terraces. \textbf{(d)} Height profile along the direction marked with black in (c).
    }
    \label{fig:z_profile}
\end{figure}

Likewise, STM images shown in the main text in Figures 1b and 2b are hereafter displayed (Figure \ref{fig:z_profile}) with height profiles. 
Both for CrCl$_3$ and CrCl$_2$ islands, the measured height is about 3 \AA, which is compatible with a Cl-Cr-Cl trilayer lying on top of a metallic substrate with reduced vdW gap.
The interaction between vdW layers and cut single crystal FCC metal surfaces, \textit{e.g.} along the (111) direction, is not purely vdW due to the presence of dangling bonds. 
The thickness profile of second layer nucleated clusters, which appear on top of the CrCl$_3$ domains in Figure \ref{fig:z_profile}a, is instead a few \AA \ higher, probably due to a larger vdW gap, similar to other monolayer materials prepared on inert/dangling bonds-free substrates  \cite{thick}.

%%%%%%%%%%%%%%%%%%%%%%%%%%%%%%%%%%%%%%%%%%%%%%%%%%%%%%%%%%%%%%%%%%%%%
%% Graphene growth
%%%%%%%%%%%%%%%%%%%%%%%%%%%%%%%%%%%%%%%%%%%%%%%%%%%%%%%%%%%%%%%%%%%%%

\section{Graphene growth}
The growth of single layer graphene (SLG) on top of the Ni(111) substrate was carried out by ethylene (C$_2$H$_4$) pyrolysis as described in the \textit{Experimental Section} of the main article. 
After the pyrolysis step, two distinct routes were taken for cooling: in the first, (\textit{i}) the sample is slowly cooled in the same ethylene partial pressure (5$\cdot$10$^{-7}$ mbar); in the second (\textit{ii}) the cooling is entirely done in ultra-high vacuum (UHV). This second process leaves residual patches of Nickel Carbide (Ni$_2$C) that need to be converted into SLG by an additional annealing step at 475$^{\circ}$C.

\begin{figure}[h]
    \centering
    \includegraphics[scale=0.4]{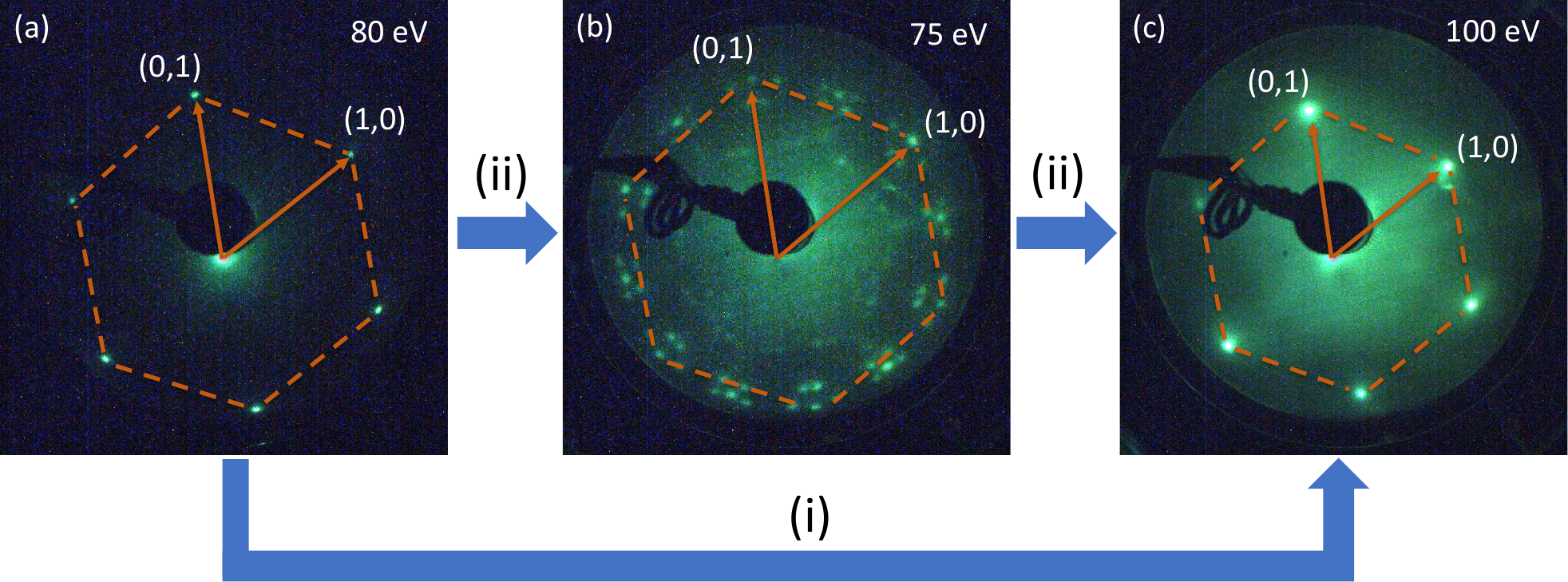}
    \renewcommand{\figurename}{\textbf{Figure}}
    \renewcommand{\thefigure}{\textbf{S\arabic{figure}}}
    \caption{
    LEED patterns of (a) Ni(111) after sputtering and annealing cycles, (b) Ni$_{2}$C/Ni(111) surface and (c) SLG/Ni(111), respectively taken at 80, 75 and 100 eV (best contrast). The dashed line hexagon (vectors) connect (point to) the Ni(111) reflections.
    }
    \label{fig:Gr_growth}
\end{figure}

Figure \ref{fig:Gr_growth} reports the LEED patterns acquired after each step during the SLG/Ni(111) substrate preparation.
Considering the case (\textit{ii}), after the sputtering and annealing cycles, the Ni(111) surface (Figure \ref{fig:Gr_growth}a) is exposed to 5$\cdot$10$^{-7}$ mbar of ethylene pressure at 575$^{\circ}$C. 
Cooling in UHV leads to a ($16\sqrt{3}\times16\sqrt{3})R30^{\circ}$-Ni$_{2}$C/Ni(111) reconstructed surface (Figure \ref{fig:Gr_growth}b) \cite{mccarroll}. 
A post annealing in UHV at 475$^{\circ}$C is necessary to convert the carbide into SLG. Since graphene has very close lattice constants to Ni(111) (2.46 \AA \ and 2.48 \AA \ respectively) \cite{batzill}, the LEED pattern is a (1$\times$1) reconstruction, where the broad overlayer spots are exactly superimposed to the substrate ones (Figure \ref{fig:Gr_growth}c).
When the graphene is cooled done in ethylene atmosphere (\textit{i}), the conversion step is not needed and the LEED appears as in Figure \ref{fig:Gr_growth}c straight after the process.

%%%%%%%%%%%%%%%%%%%%%%%%%%%%%%%%%%%%%%%%%%%%%%%%%%%%%%%%%%%%%%%%%%%%%
%% Complementary XMCD measurements on CrCl3/Au(111)
%%%%%%%%%%%%%%%%%%%%%%%%%%%%%%%%%%%%%%%%%%%%%%%%%%%%%%%%%%%%%%%%%%%%%

\section{Complementary spectroscopic characterization of CrCl$_3$/Au(111)}

\begin{figure}[h]
    \centering
    \includegraphics[scale=0.5]{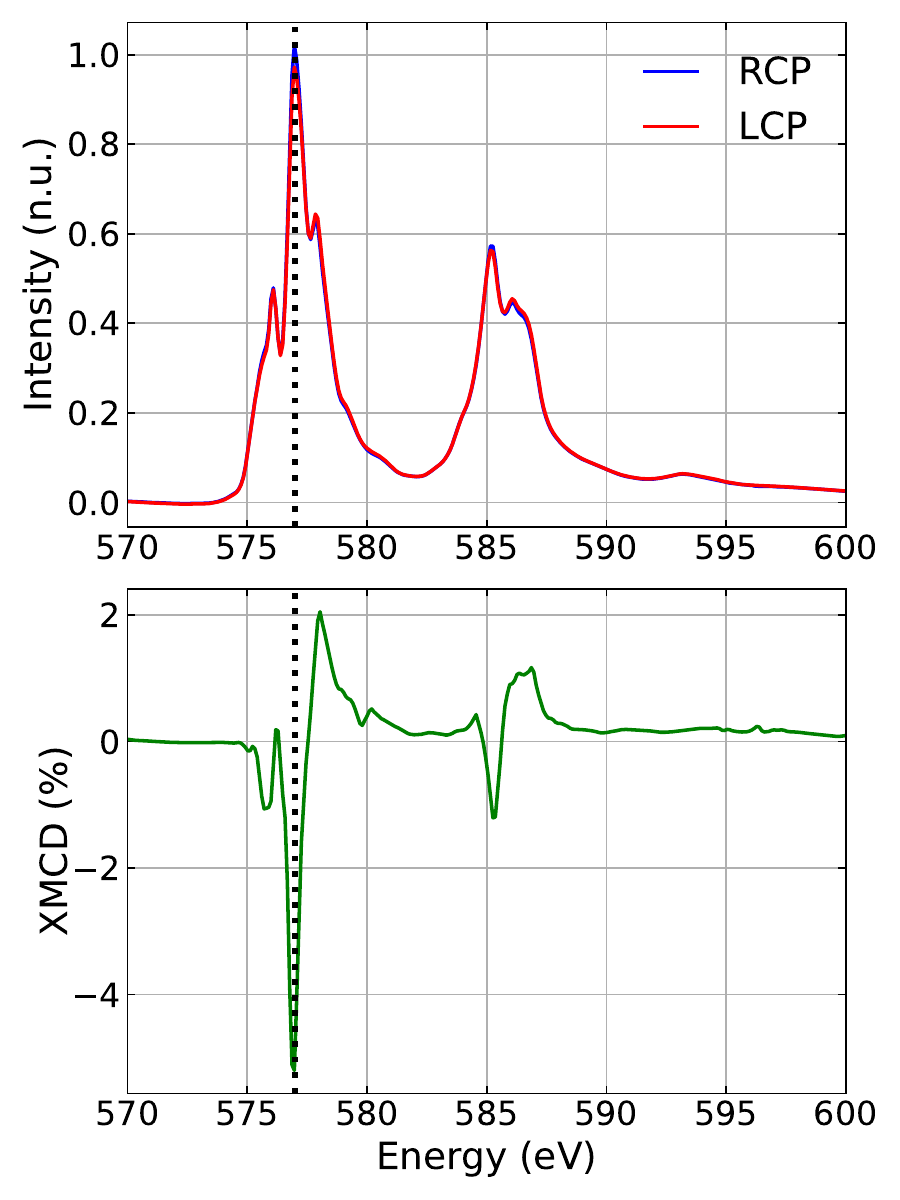}
    \renewcommand{\figurename}{\textbf{Figure}}
    \renewcommand{\thefigure}{\textbf{S\arabic{figure}}}
    \caption{
    XAS and XMCD spectra of CrCl$_3$/Au(111) taken around the Cr L$_{3,2}$ edges at 4 K, NI, and in remanence after application of 9 T magnetic field in the out-of-plane direction.
    }
    \label{fig:xmcd_Au_0T}
\end{figure}

In this section we show additional spectroscopic measurements performed on CrCl$_3$/Au(111). 
Figure \ref{fig:xmcd_Au_0T} shows XAS and XMCD spectra acquired \textit{in remanence}, at 4 K and in NI, following a previous set of measurements at 9 T magnetic field applied perpendicular to the monolayer planes. 
A finite magnetic dichroism of $\sim$6\% persists, indicating a remanent magnetic moment of Cr atoms.

\begin{figure}[h]
    \centering
    \includegraphics[scale=0.7]{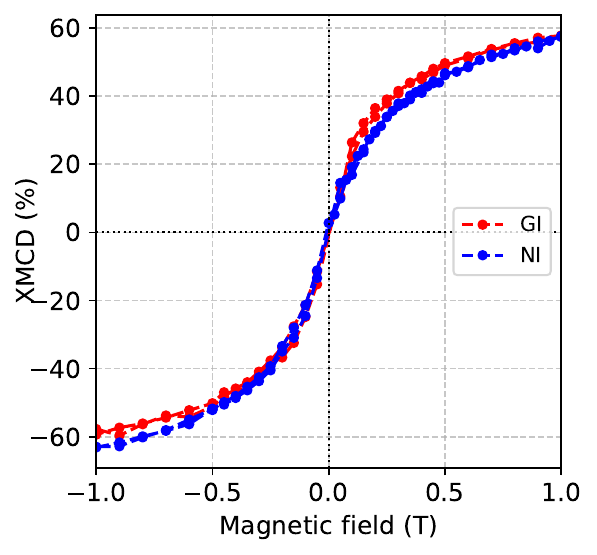}
    \renewcommand{\figurename}{\textbf{Figure}}
    \renewcommand{\thefigure}{\textbf{S\arabic{figure}}}
    \caption{
    Magnetization measurements at 4 K on CrCl$_3$/Au(111) taken with photon energy corresponding to Cr$^{3+}$ L$_{3,2}$ edges, in normal incidence (NI) and in grazing incidence (GI).
    }
    \label{fig:xmcd_Au_GI_NI}
\end{figure}

In Figure \ref{fig:xmcd_Au_GI_NI}, we compare the magnetization curves measured at normal and grazing incidence (NI vs GI) between -1 T and 1T.
No clear magnetic anisotropy is observed for CrCl$_3$/Au(111). 
No residual magnetization is measured at zero magnetic field neither, in contrast with the XMCD data. 
It should be noted that a coercive field of the order of $\sim$10 mT, as the one measured by \textit{Bedoya-Pinto et al.} for CrCl$_3$/SLG/SiC(0001) \cite{bedoya}, falls below the detection limit of the beamline superconducting magnet and it cannot be accurately measured while sweeping the field.
We conclude that XMCD and magnetization data point to a weak ferromagnetic behavior for this sample.

\begin{figure}[h]
    \centering
    \includegraphics[scale=0.7]{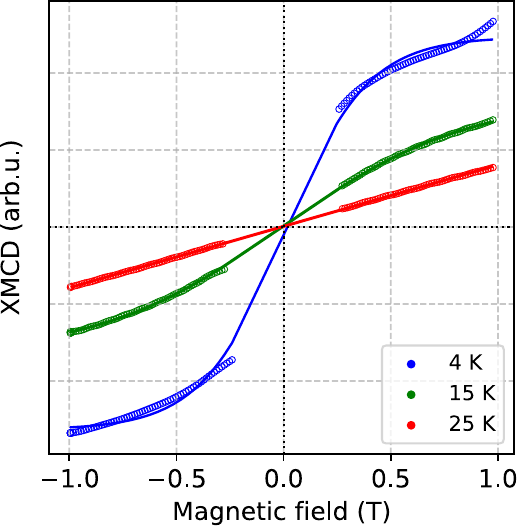}
    \renewcommand{\figurename}{\textbf{Figure}}
    \renewcommand{\thefigure}{\textbf{S\arabic{figure}}}
    \caption{Magnetization curves of CrCl$_3$/Au(111) measured at 4, 15 and 25 K, in NI. The experimental points at low magnetic field were not measured due to instability of the superconducting magnet close to zero T. Fit curves calculated with sigmoid functions are superimposed to the experimental points for better data visualization.
    }
    \label{fig:T_dep}
\end{figure}

In Figure \ref{fig:T_dep}, magnetization curves measured in NI were acquired at 4, 15 and 25 K. The measurements demonstrate that the S-shape curve at 4 K evolves rapidly towards a linear dependence by increasing the temperature, suggesting that the onset of the ferromagnetic order occurs within the 5-15 K temperature window, as in Ref. \cite{bedoya}.

%%%%%%%%%%%%%%%%%%%%%%%%%%%%%%%%%%%%%%%%%%%%%%%%%%%%%%%%%%%%%%%%%%%%%
%% XAS and XMCD simulations
%%%%%%%%%%%%%%%%%%%%%%%%%%%%%%%%%%%%%%%%%%%%%%%%%%%%%%%%%%%%%%%%%%%%%

\section{XAS and XMCD simulations}
The simulated XAS and XMCD spectra in Figures 1f and 2f in the main text were computed by ligand field multiplet (LFM) calculations within the framework of the Anderson Impurity Model, using the \textit{Quanty} software package \cite{haverkort}. 
To calculate the full multiplet structure, intra-atomic and crystal field interactions as well as Cr 3d states hybridization with the ligand valence states were considered. 
All the calculations were performed by simulating a temperature of 4 K and an external magnetic field of 9 T applied normally to the sample surface. 
The best fits of the experimental data were obtained by iterating the calculations multiple times for each relevant parameter, starting from initial guesses found in the literature for similar compounds \cite{Ghosh}.

\begin{figure}[h]
    \centering
    \includegraphics[scale=0.55]{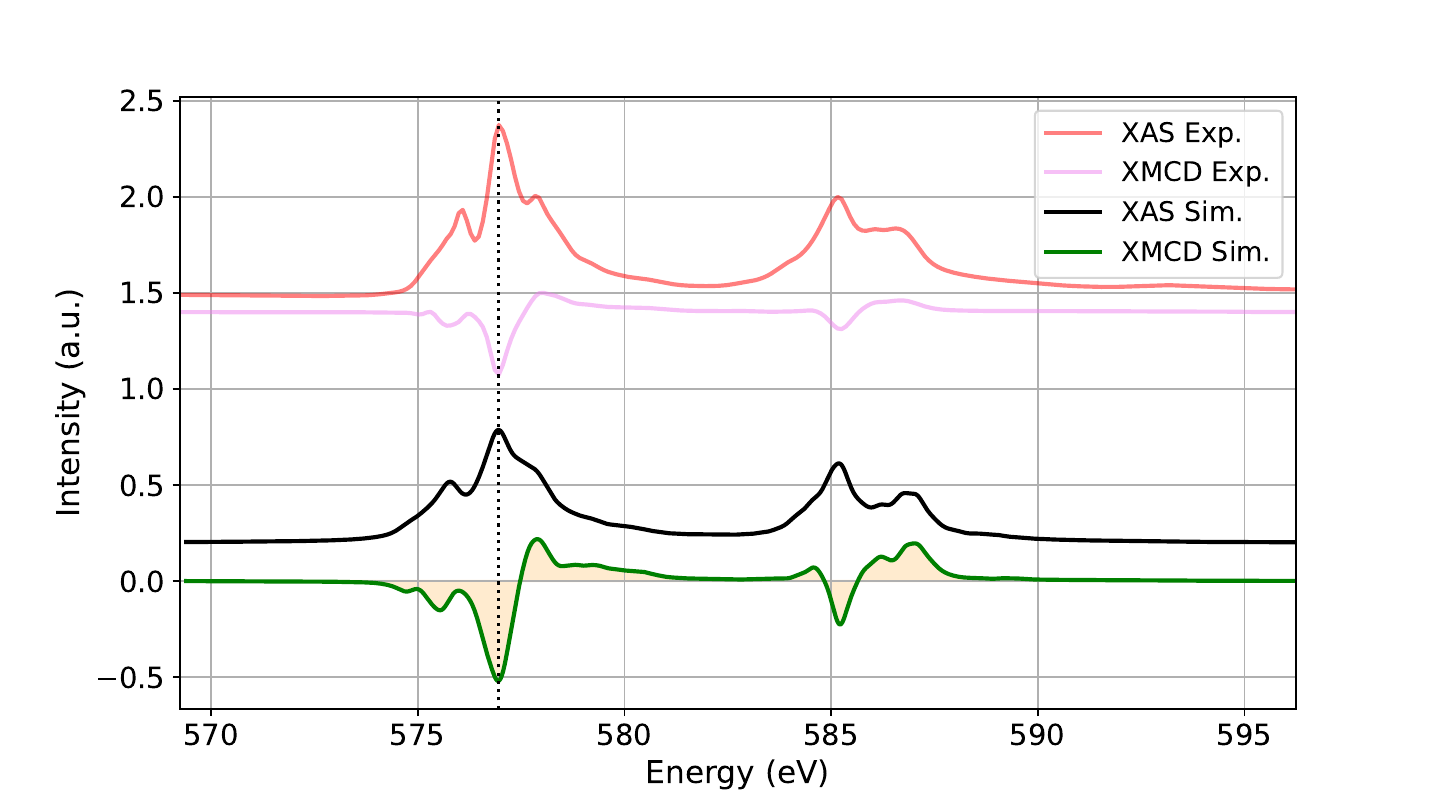}
    \renewcommand{\figurename}{\textbf{Figure}}
    \renewcommand{\thefigure}{\textbf{S\arabic{figure}}}
    \caption{
    Experimental XAS (red) and XMCD (violet) spectra of as grown CrCl$_3$ are compared with simulations (black and green respectively).
    }
    \label{fig:sim_CrCl3}
\end{figure}

For CrCl$_3$, the calculations were carried out adopting $D_{3d}$ symmetry, taking into account possible trigonal distortions of the pristine octahedral ($O_{h}$) structure. 
Besides the octahedral crystal field parameter $10Dq$, a parameter $\tau$ was also introduced to account for the splitting of the t$_{2g}$ energy levels due to trigonal distortion. 
The internal exchange field was included in the Hamiltonian via a term $H_{ex}$. 
The intra-atomic Coulomb and exchange interactions in the Hamiltonian have been calculated starting from Slater integrals reduced to 70$\%$ of their Hartree–Fock values. 
These and all the other relevant parameters and the relative best fit values are reported in the Table \ref{tab:table1}. 
The spectra are then convoluted with Gaussian (0.55 eV) and Lorentzian (0.1 eV) functions to account for the instrumental and the intrinsic broadening (Figure \ref{fig:sim_CrCl3}). 
The simulations optimally agree with the experimental curves, included the RCP and LCP XAS shown in Figures 1f and 2f. 
Notably, we found stronger crystal field interaction ($10Dq=1.75$ eV) compared to CrI$_3$ in Ref. \cite{Ghosh}, as expected due to the higher electronegativity of Cl with respect to I, and a finite t$_{2g}$ energy level splitting due to trigonal distortion ($\tau=0.20$ eV). 

The simulations allowed to estimate some noteworthy expectation values that cannot be retrieved experimentally. 
From the simulated spectra, the calculated spin angular momentum projection along the z-axis was found to be $\left< Sz \right> = 2.98 \mu_B$, consistent with the nominal value for a high spin Cr $d^3$ configuration and the value extrapolated by \textit{Bedoya Pinto et al.}  with corrected sum rules \cite{bedoya}. The orbital moment instead is essentially quenched ($\left< Lz \right> = -0.04 \mu_B$).  

\begin{table}[h]
    \centering
    \renewcommand{\tablename}{\textbf{Table}}
    \renewcommand{\thetable}{\textbf{S\arabic{table}}}
    \caption{
    \label{tab:table1}: Main parameter values used to calculate the CrCl$_3$ XAS/XMCD spectra in Figures \ref{fig:sim_CrCl3} and 1f: $d$ is the number of $3d$ electrons, $U_{dd}$ is the $3d$-$3d$ Coulomb potential, $U_{pd}$ is the $2p$-$3d$ core-hole potential, $10Dq$ and $\tau$ are the octahedral and trigonal crystal-field interactions, $V$ are the hybridization parameters for each of the given $D_{3d}$ symmetry sub-groups, $\Delta$ is the charge-transfer energy and $H_z$ is the internal exchange field. Values are expressed in eV.
    }
    \begin{tabular}{llllllllll}
        \hline \hline
        $d$ & $U_{dd}$ & $U_{pd}$ & $10Dq$ & $\tau$ & $V_{e_{g}^{\sigma}}$ & $V_{a_{1g}}$ & $V_{e_{g}^{\pi}}$ & $\Delta$ & $H_z$ \\
        \hline \hline
        3 & 4.50 & 5.50 & 1.75 & 0.20 & 1.35 & 1.25 & 1.50 & 3.00 & 0.035  \\
        \hline \hline
    \end{tabular}
    \end{table}

\begin{figure}[h]
    \centering
    \includegraphics[scale=0.55]{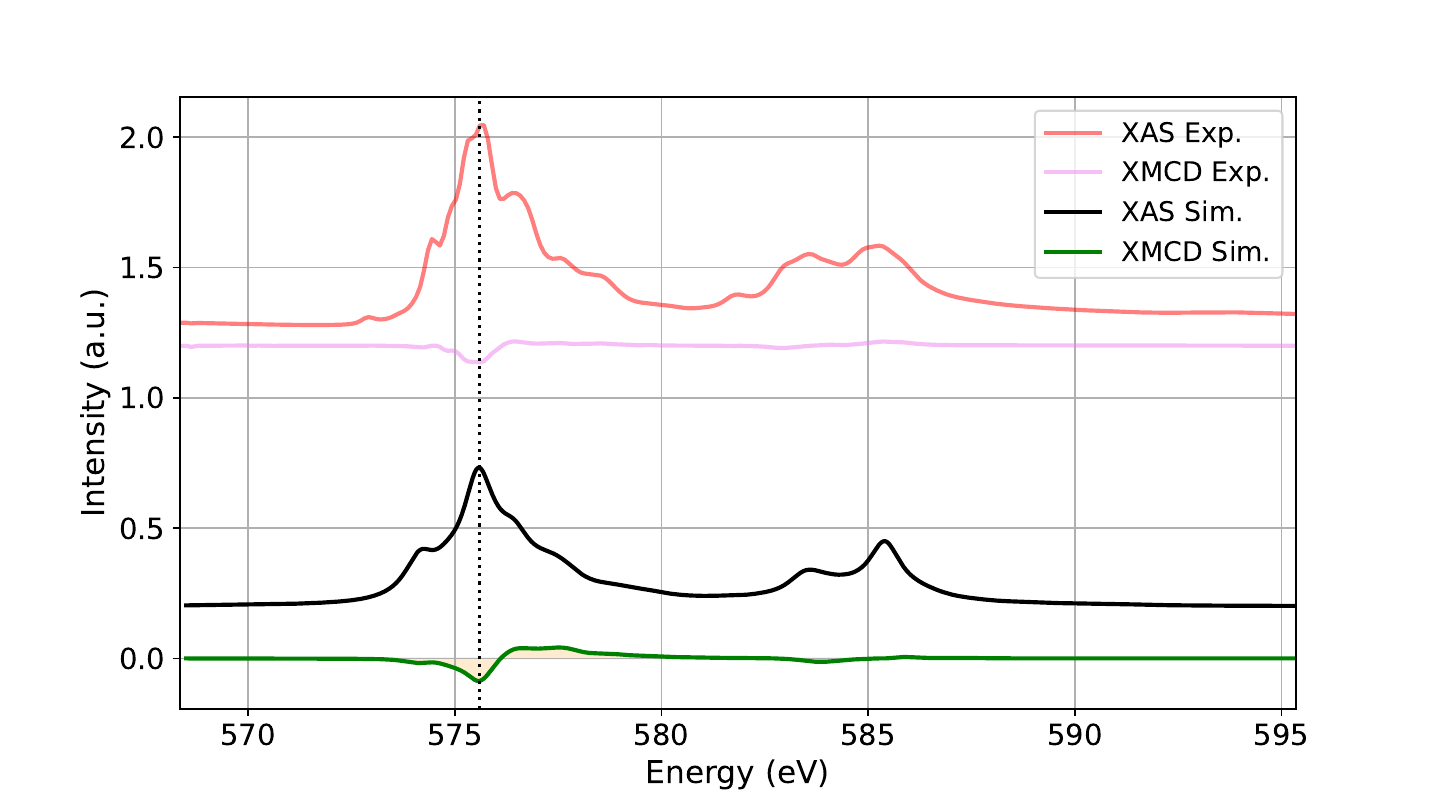}
    \renewcommand{\figurename}{\textbf{Figure}}
    \renewcommand{\thefigure}{\textbf{S\arabic{figure}}}
    \caption{Experimental XAS (red) and XMCD (violet) spectra of as grown CrCl$_2$ are compared with simulations (black and green respectively).
    }
    \label{fig:sim_CrCl2}
\end{figure}

For CrCl$_2$, the calculation were performed in octahedral symmetry, using the same trigonal $D_{3d}$ framework as for CrCl$_3$ but fixing the trigonal crystal field $\tau$ value to zero. Most of the parameters - reported in the Table \ref{tab:table2} - were kept fixed to the CrCl$_3$ values, including Slater factors and broadening. The octahedral crystal field and the charge transfer were chosen slightly smaller ($10Dq = 1.50$ eV and $\Delta=2.00$ eV). Moreover, no exchange field was included in this case in the Hamiltonian, supporting the non-ferromagnetic behavior of CrCl$_2$ as observed in the experiments. The shape of the simulated spectra as well as the XMCD magnitude are in fairly good agreement with the experimental results (Figure \ref{fig:sim_CrCl2}). The spin and orbital angular momenta were found respectively $\left< Sz \right> = -0.275 \mu_B$ and $\left< Lz \right> = 0.004 \mu_B$.

\begin{table}[h]
    \centering
    \renewcommand{\tablename}{\textbf{Table}}
    \renewcommand{\thetable}{\textbf{S\arabic{table}}}
    \caption{
    \label{tab:table2}: Main parameter values used to calculate the CrCl$_2$ XAS/XMCD spectra in Figures \ref{fig:sim_CrCl2} and 2f: $d$ is the number of $3d$ electrons, $U_{dd}$ is the $3d$-$3d$ Coulomb potential, $U_{pd}$ is the $2p$-$3d$ core-hole potential, $10Dq$ and $\tau$ are the octahedral and trigonal crystal-field interactions, $V$ are the hybridization parameters for each of the given $D_{3d}$ symmetry sub-groups, $\Delta$ is the charge-transfer energy and $H_z$ is the internal exchange field. Values are expressed in eV.
    }
    \begin{tabular}{llllllllll}
        \hline \hline
        $d$ & $U_{dd}$ & $U_{pd}$ & $10Dq$ & $\tau$ & $V_{e_{g}^{\sigma}}$ & $V_{a_{1g}}$ &  $V_{e_{g}^{\pi}}$ & $\Delta$ & $H_z$ \\
        \hline \hline
        4 & 4.50 & 5.50 & 1.50 & 0.00 & 1.35 & 1.25 & 1.50 & 2.00 & 0.00 \\
        \hline \hline
    \end{tabular}
\end{table}

\newpage

%%%%%%%%%%%%%%%%%%%%%%%%%%%%%%%%%%%%%%%%%%%%%%%%%%%%%%%%%%%%%%%%%%%%%
%% References
%%%%%%%%%%%%%%%%%%%%%%%%%%%%%%%%%%%%%%%%%%%%%%%%%%%%%%%%%%%%%%%%%%%%%

\bibliographystyle{MSP}
\bibliography{si}
\newpage

%%%%%%%%%%%%%%%%%%%%%%%%%%%%%%%%%%%%%%%%%%%%%%%%%%%%%%%%%%%%%%%%%%%%%
%%  end
%%%%%%%%%%%%%%%%%%%%%%%%%%%%%%%%%%%%%%%%%%%%%%%%%%%%%%%%%%%%%%%%%%%%%